\documentclass[aps]{revtex4}
\usepackage[dvips]{graphicx}
\usepackage{amssymb,amsmath,empheq}
\RequirePackage{lineno}

\newcommand{\cf} {{\em c.f. }}
\newcommand{\rr} {C}

\newcommand{\be} {\begin{equation}}
\newcommand{\ee} {\end{equation}}

\newcommand*\patchAmsMathEnvironmentForLineno[1]{%
  \expandafter\let\csname old#1\expandafter\endcsname\csname #1\endcsname
  \expandafter\let\csname oldend#1\expandafter\endcsname\csname end#1\endcsname
  \renewenvironment{#1}%
     {\linenomath\csname old#1\endcsname}%
     {\csname oldend#1\endcsname\endlinenomath}}%
\newcommand*\patchBothAmsMathEnvironmentsForLineno[1]{%
  \patchAmsMathEnvironmentForLineno{#1}%
  \patchAmsMathEnvironmentForLineno{#1*}}%
\AtBeginDocument{%
\patchBothAmsMathEnvironmentsForLineno{equation}%
\patchBothAmsMathEnvironmentsForLineno{align}%
\patchBothAmsMathEnvironmentsForLineno{flalign}%
\patchBothAmsMathEnvironmentsForLineno{alignat}%
\patchBothAmsMathEnvironmentsForLineno{gather}%
\patchBothAmsMathEnvironmentsForLineno{multline}%
}

\begin{document}

\title{Wormhole formation in dissolving fractures}

\author{P. Szymczak}
\affiliation{Institute of Theoretical Physics, Warsaw University,
Ho\.{z}a 69, 00-618, Warsaw, Poland}
\email{piotrek@fuw.edu.pl}
\homepage{http://www.fuw.edu.pl/~piotrek}

\author{A. J. C. Ladd}
\affiliation{Chemical Engineering Department, University of Florida,
Gainesville, FL  32611-6005, USA}
\email{tladd@che.ufl.edu}
\homepage{http://ladd.che.ufl.edu/}

\begin{abstract}
We investigate the dissolution of artificial fractures with
three-dimensional, pore-scale numerical simulations. The fluid velocity in
the fracture space was determined from a lattice-Boltzmann
method, and a stochastic solver was used for the transport of
dissolved species. Numerical simulations were used to study
conditions under which long conduits (wormholes) form in an 
initially rough but spatially homogeneous fracture. 
The effects of flow rate, mineral dissolution rate and 
geometrical properties of the fracture were investigated, and 
the optimal conditions for wormhole formation determined. 
\end{abstract}

\maketitle

\section{Introduction}

A number of experimental and numerical studies  of dissolution in fractured or porous rock have established that the evolving topography of the pore space depends strongly on the fluid flow and mineral dissolution rates.
Remarkably, there exists a parameter range in which positive feedback between fluid
transport and mineral dissolution leads to the spontaneous formation of pronounced
channels, frequently referred to as ``wormholes''. Spontaneous channeling of a reactive front has been shown to be important for a number of geophysical processes, such as diagenesis~\citep{Chen1990,Boudreau1996}, melt migration~\citep{Daines1994,Aharonov1995,Kelemen1995,Spiegelman2001}, terra rosa formation~\citep{Merino2008}, development of limestone caves~\citep{Gro95,Han98}, and sinkhole formation by salt dissolution~\citep{Shalev2006}. Further details can be found in review articles and books on reactive transport and geochemical self-organization,
~\citep[e.g.][]{Ste90,Ortoleva1994,MacQuarrie2005,Steefel2005,Steefel2007}.

Wormholes play an important role in a number of geochemical applications, most notably
${\rm CO}_2$ sequestration~\citep{Cailly2005,Kang2006,King2007}, risk assessment of contaminant migration in groundwater~\citep{Fryar1998} and stimulation of petroleum reservoirs~\citep{Eco2000,Kalfayan2000}. Selecting the optimal flow rate is an important issue in reservoir stimulation, so as to achieve the maximum increase in permeability for
a given amount of reactant~\citep{Fredd1998,Golfier2002,Panga2005,Kalia2007,Cohen2008}.
If the acid is injected too slowly, significant dissolution occurs only
at the inlet, and the permeability of the system remains almost unchanged.
At the other extreme of high injection velocities dissolution tends to be uniform throughout the sample. However, the increase in permeability is again insignificant, since the reactant is consumed more or less uniformly throughout the fracture, making only an incremental change to the permeability. Moreover, some of the reactant may escape unused. The most efficient stimulation is obtained for intermediate injection rates, 
where the reactive flow self-organizes into a small number of distinct channels, while the rest of the medium is effectively bypassed. This focusing mechanism leads to much more efficient use of reactant, since the development of channels causes a large increase in permeability with a relatively small consumption of reactant.

Experimental studies of wormhole formation have used a 
variety of porous systems; plaster dissolved by water~\citep{Daccord1987,Dac87}, 
limestone cores treated with hydrochloric acid~\citep{Hoe88} and salt-packs dissolved with under-saturated salt solution~\citep{Kelemen1995,Golfier2002}. Recently, a variety of dissolution patterns in single rock fractures have been reported~\citep{Dur01,Dij02,Gouze2003,Det03,Polak2004,Detwiler2008}, depending on the chemical and physical characteristics of the fracture-fluid system. The physico-chemical mechanisms behind the pattern formation are not yet understood in detail. Linear stability analysis has been used to investigate the conditions required for the break up of a planar dissolution front~\citep{Chadam1986,Ort87,Hinch1990}, but these results only pertain to the initial stages of channel formation, where the front perturbations are small. 
The later stages of channel evolution are strongly nonlinear and here numerical methods are needed.
The numerical models used to study wormholing in porous media
fall into four broad categories. (1) Single wormhole models~\citep{Hung1989,Buijse2000}, 
in which a channel of a given shape is created in the porous matrix. The reactant concentration field inside the wormhole is determined and the growth velocity calculated. (2) Darcy-scale models~\citep{Golfier2002,Panga2005,Kalia2007} based on continuum equations with effective variables such as dispersion coefficients, Darcy velocity and bulk reactant concentrations. These average variables replace the microscopic diffusion constant, fluid velocity, and reactant concentration. (3) Network models~\citep{Hoe88,Fredd1998}, which
model fluid flow and dissolution in a network of interconnected pipes, where the diameter of each network segment or pipe is increased in proportion to the local reactant consumption. (4) Pore-scale numerical simulations~\citep{Bek95,Kang2002,Kang2003,Kang2006b}. In these calculations the equations for fluid flow, reactant transport and chemical kinetics are solved in an explicitly three-dimensional pore space. Although computationally intensive such models provide detailed information on the evolution of the fluid velocity, reactant concentration and topography without invoking effective parameters such as mass transfer coefficients. This is the approach followed in the present work, however in the context of fracture dissolution.

In studies of fracture dissolution, and particularly in theoretical 
investigations of cave formation, a one-dimensional model of a single fracture is frequently used 
\citep[e.g.][]{Drey90,Gro94,Dre96,Dij98}. The fracture aperture
(the distance between the rock surfaces) is assumed to depend on a single spatial variable, the distance from the inlet.  
Although analytically tractable, one-dimensional models cannot account for wormhole formation, and thus they are only relevant at the extremes of high and low flow rate where the dissolution is expected to be uniform. 
In two-dimensional models of dissolving fractures~\citep{Han98,Che02,Detwiler2007}, the fluid velocity and reactant concentration are averaged over the aperture
of the fracture. The key simplifications are the Reynolds (or lubrication) approximation for  the fluid velocity~\citep{Adl00} and the use of effective reaction rates. These models are technically similar to Darcy-scale models, with the local permeability determined by the aperture; they produce realistic-looking erosion patterns and correlate positively with experimental results~\citep{Detwiler2007}. However the Reynolds approximation may significantly overestimate the flow rate~\citep{Bro95,Oro98,Nic99}, especially for fractures of high roughness and small apertures. Moreover, under certain geological and hydrological conditions, large pore-scale concentration gradients develop and in such cases volume averaging can introduce significant errors~\citep{Li2007a,Li2008},  sometimes not even capturing the correct reaction direction.

The most fundamental approach is to directly solve 
equations for fluid flow, reactant transport, and chemical kinetics within the 
fracture space. This approach was pioneered by~\citet{Bek97}, who solved
the flow and transport equations using finite-difference schemes. Better resolution
is offered by lattice Boltzmann methods, which have been used in dissolution simulations at the pore scale \citep{Ver02,Kang2002,Kang2003,Szy04b,Szy06,Verhaeghe2006,Kang2006b,Arnout2008} and at the Darcy-scale \citep{OBrien2002,OBrien2003}.
Here we combine velocity field calculations from an implicit lattice-Boltzmann
method~\citep{Ver99} with a transport solver based on random walk algorithms that incorporates the chemical kinetics at the solid surfaces~\citep{Szy04}.
Advances in numerical algorithms for flow and transport allow us to simulate 
systems of relevance to laboratory experiments without resorting to semi-empirical approximations~\citep{Szy04b}. In the simulations reported here, as many as 50 interacting wormholes have been studied (\cf Fig.~\ref{dissol}),
comparable to systems modeled by state of the art Darcy-scale simulations~\citep{Cohen2008}, while maintaining pore-scale resolution.

We have investigated wormhole formation in a simple artificial geometry, where one of the fracture surfaces is initially flat, and the other is textured with several thousand randomly placed obstacles.
The geometry is similar to that studied experimentally by Detwiler et al.~\citep{Det03,Detwiler2007} and has the advantage that it shows no discernible long-range spatial order. The correlation length is of the order of the distance
between the obstacles, and is much smaller than the system size. Although it lacks
the self similarity of natural fractures, it provides a useful starting point
for numerical analysis of wormholing, since the initial structure contains no nascent channels.

The aim of this paper is to discover the range of conditions under which long conduits form
in a initially rough but spatially homogeneous fracture. Dissolution was studied under conditions corresponding to a constant pressure drop across the sample and to a constant flow rate. Constant pressure drop is representative of the early stages of karstification~\citep{Drey90}, whereas constant flow rate is more relevant for reservoir stimulation~\citep{Eco2000}. In this context, we have numerically determined the conditions needed to maximize the permeability increase for a given amount of reactant.

We investigate the effects of flow rate (characterized by the P\'eclet number), mineral dissolution rate (characterized by the  Damk\"ohler number), and geometrical properties of the fracture. 
The P\'eclet number measures the relative magnitude of convective and diffusive transport of the
solute, 
\begin{equation}
Pe = {\bar v} {\bar h}/D,
\label{Pec}
\end{equation}
where ${\bar v}$ is a mean fluid velocity, ${\bar h}$ is the mean
aperture and $D$ is the solute diffusion coefficient. In this work, ${\bar v} = Q/(W {\bar h})$ is related to the volumetric
flow rate, $Q$, and the the mean cross sectional area, $W{\bar h}$, where $W$ is the width of the
fracture. The Damk\"ohler number, 
\begin{equation}
Da=k/{\bar v},
\label{Dam}
\end{equation}
relates the surface reaction rate to the mean fluid velocity.
The relevant geometric characteristics are harder to quantify. \citet{Han98}
argued that the key geometrical factor determining the intensity of wormholing is
the statistical variance of the aperture field, $\sigma$, relative to the mean aperture,
$f = \sigma/\bar h$.
Here, we present numerical evidence that the total extent of contacts between the surfaces may play an important role as well. These contacts need not be load bearing;
the dynamics remains qualitatively the same if the surfaces are sufficiently close
that the local fluid flow is strongly hindered. The transition between uniform dissolution and
channeling seems to occur rather sharply in our simulations, at the point where the contact
regions make up $5 \%-10\%$ of the total fracture area.

\section{Numerical model}

\begin{figure}[t]
\center\includegraphics[width=20pc]{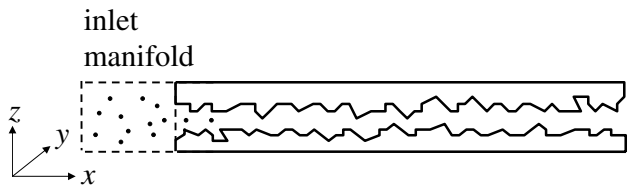}
\caption{The geometry of the experiment: a corrugated glass surface (upper) is
matched with a soluble flat plate (lower). The plates are held in a
fixed position and reactant flows from an inlet manifold designed to produce a uniform
concentration and flow field at the inlet.}\label{geometry}
\end{figure}

To investigate channel growth and interaction in a dissolving
fracture, we use a pore-scale numerical model
\citep{Szy04b} in which the fracture space is defined by two-dimensional
height profiles $h^u(x,y)$ and $h^l(x,y)$ representing the upper and lower
fracture surfaces. The velocity field in the fracture space
is calculated by an implicit lattice-Boltzmann technique
\citep{Ver99}, while the transport of dissolved species is modeled
by a random walk algorithm, which efficiently incorporates the
chemical kinetics at the solid surfaces~\citep{Szy04}. The
fracture surfaces are discretized into pixels and the
height of each pixel is eroded in response to contacts by tracer
particles;  for the results reported below either $200 \times 400$, $400 \times 400$,
or $800 \times 800$  pixels were used. The time evolution of the
velocity field and the local aperture variation in the fracture are
determined by removing small amounts of material at each step, and
recalculating the flow field and reactant fluxes for the
updated topography.

\subsection{Flow field calculation}

In laboratory-scale fractures, the Reynolds number is less than 1~\citep{Dur01,Dij02,Det03};
it is also small during the initial stages of cave formation~\citep{Palmer1991,Gro94}.
Thus, inertia can reasonably be neglected, and fluid motion is then governed
by the Stokes equations
\begin{equation} \label{Stokes}
\nabla \cdot {\mathbf{v}}=0; \; \; \; \eta \nabla^2 {\mathbf{v}}=\nabla p,
\end{equation}
where ${\mathbf{v}}$ is the fluid velocity, $\eta$ is the viscosity
and $p$ is the pressure. The velocity field in the fracture has been calculated
using the lattice-Boltzmann method with ``continuous bounce-back'' rules
applied at the solid-fluid boundaries~\citep{Ver00}. The accuracy of these boundary
conditions is insensitive to the position of the interface with respect to the lattice,
which allows the solid surface to be resolved on length scales less than a grid spacing;
thus the fracture surfaces erode smoothly. It has been shown~\citep{Ver02} that the
flow fields in rough fractures can be calculated with one-half to one-quarter
the linear resolution of the ``bounce-back'' boundary condition, leading to
an order of magnitude reduction in memory and computation time. A further order
of magnitude saving in computation time can be achieved by a direct solution of the
time-independent lattice-Boltzmann model~\citep{Ver99}, rather than by time
stepping. These improvements have previously allowed us to calculate velocity
fields in laboratory-scale fractures~\citep{Szy04b}. The calculation of the flow field
in a $200 \times 400$ fracture takes about 1 minute at the beginning of the dissolution
process, and up to 15 minutes during the final stages of dissolution. The processor was a single core of an Intel Pentium P4D clocked at 3GHz.

\subsection{Solute transport modeling}\label{model}

Solute transport in the fracture is modeled by a random walk algorithm that
takes explicit account of chemical reactions at the pore surfaces. The concentration
field is represented by a distribution of tracer particles, each representing $n$ solute molecules. We use a standard stochastic solution  to the convection-diffusion equation,
\citep{Klo93,Hon93}
\begin{equation}\label{CDeqn}
\partial_t c + {\bf v \cdot \nabla} c = D \nabla^2 c.
\end{equation}
in which individual particles are tracked in space and time,
\begin{equation}
{\bf r}_i(t+\delta t)= {\bf r}_i(t)+ {\bf v}({\bf r}_i) \delta t + \sqrt{2 D \delta t} {\bf
\Gamma}.
\label{class}
\end{equation}
The flow field, ${\bf v}({\bf r})$, is derived from the implicit
lattice-Boltzmann simulation and
${\bf \Gamma}$ is a Gaussian random variable of zero mean and unit variance.
The fluid velocity at the particle position is interpolated from the surrounding grid
points, while the time step $\delta t$ is chosen
such that the displacement in one step is smaller than $0.1 \delta x$, where
$\delta x$ is the grid spacing. To account for chemical erosion at the
fracture surfaces, we calculate the dissolution flux at each boundary pixel, assuming
a first-order surface reaction
\begin{equation}
J_{\perp} = k (c_s-c_0),
\end{equation}
where $c_s$ is the saturation concentration, $c_0$ is the local concentration at the surface
and $k$ is the surface reaction rate. The solute flux is normal to the surface, and forms a boundary condition to the transport solver,
\begin{equation}
- D (\nabla_{\perp} c)|_0 = {\bf n} k(c_s -c_0),
\label{jj}
\end{equation}
where ${\bf n}$ points into the fluid, and $\nabla_{\perp} = {\bf n} {\bf n} \cdot \nabla$. The notation $\nabla(\ldots)|_0$ indicates a gradient at the surface.

It is convenient to introduce a reactant concentration field $\rr$, so as to simplify the boundary condition in Eq.~\eqref{jj}; in the present context 
\begin{equation}
\rr=c_s-c
\end{equation}
is the undersaturation, measuring the deviation of $c$ from the saturation concentration. The boundary condition \eqref{jj} is then
\begin{equation}\label{eq:bc}
D (\nabla_{\perp} \rr)|_0 = {\bf n} k \rr_0
\end{equation}
In a different situation, for instance dissolution of a fracture by a strong acid, the reactant field, $\rr$, is the acid concentration. It is most convenient to define $\rr$ so that the boundary condition always takes the form of Eq.~\eqref{eq:bc}; the convection-diffusion equation \eqref{CDeqn} is the same in both cases.

The drawback of the classical random walk method~\citep{Bek95} is that a very large number
of particles must be tracked simultaneously, so that the concentration near the
rock surface can be determined accurately enough to obtain a statistically
meaningful dissolution flux. However, there is a considerable simplification
in the case of linear dissolution kinetics, where it is possible to derive a single-particle
stochastic propagator that satisfies the boundary condition in Eq.~\eqref{eq:bc} ~\citep{Szy04}.
The diffusive part of the particle displacement in the direction perpendicular to the
fracture surface ($z$) is then sampled from the distribution
\begin{eqnarray}
G_d(z,z^\prime,\delta t) &=&  \frac{1}{\sqrt{4 \pi D \delta t}} \left( e^{-(z-z^\prime)^2/4D \delta t} + e^{-(z+z^\prime)^2/4D \delta t} \right) \nonumber \\ &-& \frac{k}{D} e^{k (k \delta t + z + z^\prime)/D} \mbox{Erfc} \left( \frac{z+z^\prime+2k \delta t}{\sqrt{4D \delta t}}\right)
\label{prop}
\end{eqnarray}
In Eq.~\eqref{prop}, $z^\prime$ and $z$ are the distances of the tracer particle from the surface at the beginning and at the end of the time step respectively. The boundary condition
\eqref{eq:bc} implies that the amount of reactant represented by a single tracer particle $n(t)$ decreases in time according to
\begin{equation}
n(t+\delta t)=n(t) \int_0^\infty G_d(z,z^\prime,\delta t) dz.
\end{equation}
The integral can be calculated analytically,
\begin{eqnarray}
\hspace{1em} n(t+\delta t)/n(t) &=& e^{k(z^\prime+k \delta t)/D} \mbox{Erfc}\left(\frac{z^\prime+2k\delta t}{\sqrt{4 D \delta t}}\right) \\ &+& \mbox{Erf} \left(-\frac{z^\prime}{\sqrt{4 D \delta t}}\right), \nonumber
\label{conc} 
\end{eqnarray}
and for $k>0$ the amount of material represented by the tracer is reduced,
\begin{equation}
n(t+\delta t)/n(t) < 1.
\end{equation}
In order to apply Eq.~\eqref{prop} to a complex topography, the timestep $\delta t$ must be limited, so that within each step a particle only samples a small portion of the fracture surface, which then appears planar. In our simulations $\sqrt{D \delta t} \sim 10^{-2} {\bar h}_0$, where ${\bar h}_0$ is the initial mean fracture aperture.

The side walls of the fracture (parallel to the flow direction)
are solid and inert; thus reflecting boundary conditions for the solute transport (Eq.~\ref{jj} with $k=0$)
are imposed at $y=0$ and $y = W$. At the fracture inlet ($x = 0$),
a reservoir boundary condition of constant concentration $\rr = \rr_{in}$ is applied, while a saturation
condition $\rr = \rr_{out} = 0$  is assumed at the outlet boundary, $x = L$. These boundary conditions are implemented according to the algorithms described in~\citep{Szy03}, which contain a number of subtleties.
Because mineral concentrations in the solid phase
are typically much larger than reactant concentrations in the aqueous phase,
there is a large time-scale separation between the relaxation of the concentration
field and the evolution of the fracture topography. We therefore make a quasi-static approximation,
solving for the time-independent velocity and concentration fields in a fixed fracture geometry. The quasi-static approximation may break down in cases where the reactant is much more concentrated and the reaction kinetics are fast; acid erosion by HCl is a possible example of this.

The steady dissolution flux in the fracture can be calculated by tracking individual tracers, using the following algorithm \citep{Szy03,Szy04}:

\begin{enumerate}

\item Sample the initial position of a tracer particle within the inlet manifold indicated in
Fig.~\ref{geometry}. Assign the initial number of reactive molecules represented by a
tracer,
\begin{equation}
n(0) = \frac{V_{0}}{N_{tot}} {\rr}_{in}:
\end{equation}
$V_0$ is the volume of the inlet manifold, $N_{tot}$ is the total number of particles
to be sampled, and ${\rr}_{in}$ is the inlet concentration of reactant.

\item Propagate the particle for a single time step $\delta t$, according to
Eq.~\eqref{class}. If it comes within a cutoff value $z_c$, of any surface element,
then sample the diffusive part of the particle displacement perpendicular to the wall
from Eq.~\eqref{prop}, change $n(t)$ according to Eq.~\eqref{conc}, and 
increment the dissolution flux counter at the surface element ($i$) closest to the particle by
\begin{equation}
\Delta J_i = \frac{n(t+\delta t)-n(t)}{S_i \delta t},
\end{equation}
where $S_i$ is the area of the surface element.

\item The random walk described by repeating step 2 many times is terminated when $n(t)/n(0)$ falls below a preset threshold, or when the particle leaves the system through the inlet or outlet. Random walks are also terminated when the particle fails to enter the fracture at the first step, but these must be counted, even though they do not contribute to the erosion flux.

\item Upon completion of $N_{tot}$ random walks, remove
material from the fracture walls in proportion to the accumulated fluxes, $J_i$.
The total amount of material is chosen to be sufficiently small that the evolution
of fracture topography appears continuous. 

\item The cutoff distance was set to $z_c = 10 \sqrt{2 D \delta t}$, $N_{tot}$ was of the order of $10^6$, and the threshold below which a tracer is deleted was $10^{-6}n(0)$.

\end{enumerate}

The above scheme can be used to calculate concentration profiles in large
fractures, since it is more computationally efficient than a
typical stochastic algorithm where the local concentration field is needed to
determine the dissolution flux. In that case the number of tracer particles ($N_{tot}$) required for statistically significant erosion rates is several orders of magnitude larger.

The time evolution of the velocity field and local
aperture are determined by iteration, removing small amounts of material at each step.
The dissolution-induced aperture change in the fracture over the time, $\Delta \bar{h}$, is
related to the mean dissolution flux ${\bar J}=\sum_i J_i S_i/\sum_i S_i$ by
\begin{equation}
\Delta {\bar h} = \frac{{\bar J} \Delta t}{c_{sol}} \frac{\nu_{sol}}{\nu_{aq}}
\end{equation}
where $c_{sol}$ is the concentration of the solid component and $\nu_{aq}$,
$\nu_{sol}$ are the stoichiometric numbers of the aqueous and solid species.
It is more computationally efficient to keep $\Delta \bar{h}$ constant in each erosion cycle,
and then increment the time accordingly,
\begin{equation}
\Delta t = \frac{c_{sol} \Delta {\bar h}}{\bar J} \frac{\nu_{aq}}{\nu_{sol}}.
\label{time0}
\end{equation}
In the simulations reported here, $\Delta \bar{h} = 0.01 {\bar h}_0$. Test calculations with smaller values of  $\Delta \bar{h} $ (down to $\Delta \bar{h} = 0.001 {\bar h}_0$) confirmed that the patterns are insensitive to the magnitude of the erosion step in this parameter range.

We will use a dimensionless timescale, based on the time for high flow rate dissolution in a parallel channel. In this idealized system, the reactant concentration is everywhere uniform and equal to the inlet concentration $C_{in}$. The spacing between the plates is set to ${\bar h}_0$, the initial value of the mean fracture aperture, and the reference reaction rate is chosen so that the product of P\'eclet and Damk\"ohler numbers is unity, $Pe Da = k {\bar h}_0/D = 1$. We define the characteristic time $\tau$ for a plate to erode by ${\bar h}_0$:
\begin{equation}
\tau = \frac{\bar{h}_0^2}{D} \frac{c_{sol}}{C_{in}} \frac{\nu_{aq}}{\nu_{sol}}.
\label{tau}
\end{equation}
Equation \eqref{time0} can then be rewritten in terms of a dimensionless time $\Delta t/\tau$,
\begin{equation}
\frac{\Delta t}{\tau} = \frac{D C_{in}}{\bar{h}_0^2} \frac{\Delta \bar{h}}{\bar J}
\label{time1}
\end{equation}

It is important to stress that the above model contains no
free parameters or effective mass-transfer coefficients. Instead the fundamental
equations for fluid flow, reactant transport, and chemical kinetics are
solved directly. The simulations incorporate the explicit topography of the
pore space, and the transport coefficients--viscosity, diffusivity, and
reaction rate--are determined independently.

\subsection{Validation of the numerical model}~\label{kdp}

The numerical model has been validated~\citep{Szy04b} by comparison with experimental data
obtained with an identical initial topography~\citep{Det03}. The experimental system was
created by mating a $99 \times
152 {\rm mm}$ plate of textured glass (spatial correlation length of $\sim 0.8
{\rm mm}$) with a flat, transparent plate of potassium-dihydrogen-phosphate
(KDP). The relative position of the two surfaces was
fixed during the experiment, eliminating the effects of confining pressure,
which are hard to control experimentally~\citep{Dur01} and even harder to model
numerically~\citep{Ver02}. The fracture was dissolved by an
inflowing solution of KDP at $5\%$ undersaturation. High spatial resolution data
($1192 \times 1837$, $0.083 \times 0.083 {\rm mm}$ pixels) was obtained for the evolution
of the local fracture aperture as a function of spatial position~\citep{Det03}.
The experiments were conducted at two different hydraulic gradients, 
corresponding to initial mean velocities $\bar{v}=0.029 \, \mbox{cm s}^{-1}$ and $0.116 \, \mbox{cm s}^{-1}$.
The other parameters characterizing the system are the diffusion coefficient of KDP in water, $D=6.8 \times 10^{-6}\,  \mbox{cm}^2 \, \mbox{s}^{-1}$, the initial mean aperture, $\bar{h}_0=0.0126 \, \mbox{cm}$, and the reaction
rate, $k=5.2 \times 10^{-4} \, \mbox{cm s}^{-1}$. Note that in~\cite{Szy04b} the reaction rate was erroneously reported to be smaller by a factor of two, $k=2.6 \times 10^{-4} \mbox{cm s}^{-1}$; however the correct value was used in all of the calculations reported there.
The P\'eclet and Damk\"ohler numbers calculated for these parameters are, from Eqs.~\eqref{Pec} and~\eqref{Dam}, $Pe=54$, $Da=0.018$ ($\bar{v}=0.029 \, \mbox{cm s}^{-1}$) and $Pe=216$, $Da=0.0045$ ($\bar{v}=0.116 \, \mbox{cm s}^{-1}$).

\begin{figure*}
\center\includegraphics[width=36pc]{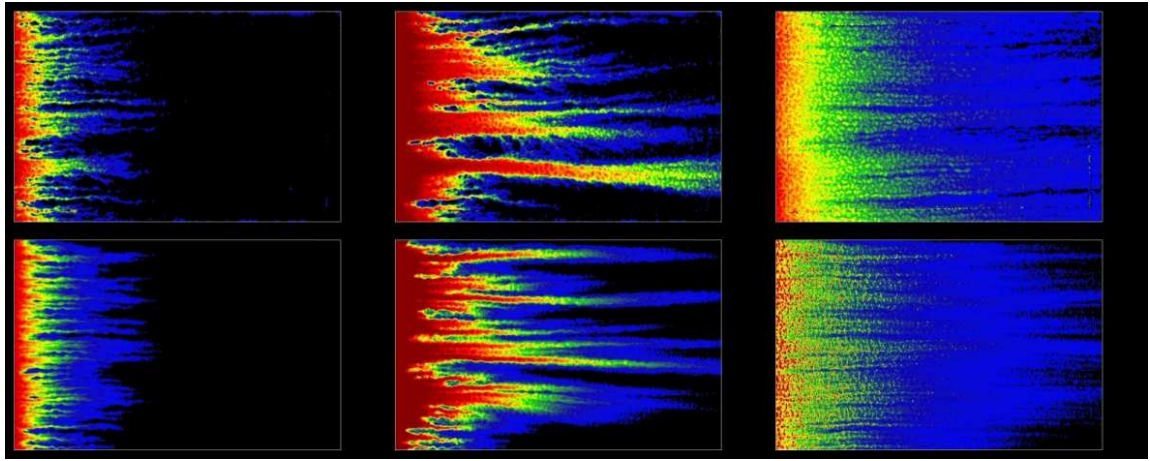}
\caption{(Color online)
Erosion of the lower surface (initially flat) during dissolution of a laboratory-scale fracture. Dissolution
patterns for $Pe=54, Da=0.018$ are shown at $\Delta \bar{h}= \bar{h}_0/2$ in the left panels
and at $\Delta \bar{h}=\bar{h}_0$ in the middle panels; the right panels show dissolution patterns
at $Pe=216, Da=0.0045$, $\Delta \bar{h}= \bar{h}_0$. The simulations are shown in the lower panels and the corresponding experimental results are shown in the upper panels. The lightest/red shading indicates the deepest erosion while dark gray/blue indicates the least erosion; the intermediate colors are yellow (higher) and green (lower). The flow direction is from left to right.}\label{tails2}
\end{figure*}

Figure~\ref{tails2} shows a comparison of the dissolution patterns obtained by simulation 
and experiment; the initial topographies in the simulation and the experiment were the same. The contour levels in the figure represent the change in height of the lower
(dissolving) surface (Fig.~\ref{geometry}) as a function of time
\begin{equation}
\Delta h(x,y;t) = h^l(x,y,0)-h^l(x,y,t).
\end{equation}
At higher flow rates, unsaturated fluid penetrates deep inside the fracture and dissolution
tends to be uniform throughout the sample (right panels in Fig.~\ref{tails2}), while at the lower flow
rate erosion is slower and inhomogeneous (middle and left panels in Fig.~\ref{tails2}).
The dissolution front is unstable to fingering~\citep{Ort87}, since an increase in permeability
within a channel enhances solute transport through it, reinforcing its growth. As dissolution proceeds, the
channels compete for the flow and the growth of the shorter channels eventually ceases.
At the end of the experiment, the flow is focused in a few main channels, while
most of the pore space is bypassed.

The experimental and numerical dissolution patterns are similar. At
low P\'eclet number, the dominant channels (Fig.~\ref{tails2}) develop at the same
locations in the simulation and experiment, despite the strongly nonlinear
nature of the dissolution front instability. While there are differences in the
length of the channels, relatively small changes (of the order of 10\%) in the
diffusion constant, $D$, or rate constant, $k$, can lead to comparable
differences in the erosion patterns. Our results suggest that the simulations are capturing the effects of the complex topography of the pore space; a more extensive and quantitative discussion, including histograms of aperture distributions at different P\'eclet numbers, can be found in~\cite{Szy04b}.

\section{Artificial fracture geometries}\label{art}

The computational model described in Sec.~\ref{model} was used to simulate dissolution in
artificial fractures, with numerically generated topographies.
Initially, the lower surface of the fracture is flat, while the upper surface
is textured with several thousand identical
cubical protrusions (parallelepipeds of height $2 \delta x$ and base $3 \delta x \times 3 \delta x$, where
$ \delta x$ is the pixel size). The protrusions were placed on a square lattice and then randomly shifted by $\pm  \delta x$ in the lateral ($y$) direction, which eliminates all the
straight flow paths between the inlet and outlet. The resulting fracture
has short range spatial correlations, and no discernable
long-range structure. The fracture geometry can be characterized statistically by the fractional coverage of
protrusions, $\zeta$. If the obstacles span the entire height of the fracture
aperture, the initial geometry has a relative roughness
\begin{equation}
f=\frac{\sigma}{\bar{h}} = \sqrt{\frac{\zeta}{1-\zeta}}
\end{equation}
where $\sigma$ is the variance of the aperture field. Most of the simulations discussed here have been carried out for $\zeta=0.5$,
which corresponds to a relatively rough fracture $f=1$; however we also investigated smoother fractures,
with $\zeta$ as small as 0.025 (see Sec.~\ref{geom}).

\begin{figure}
\center\includegraphics[width=28pc]{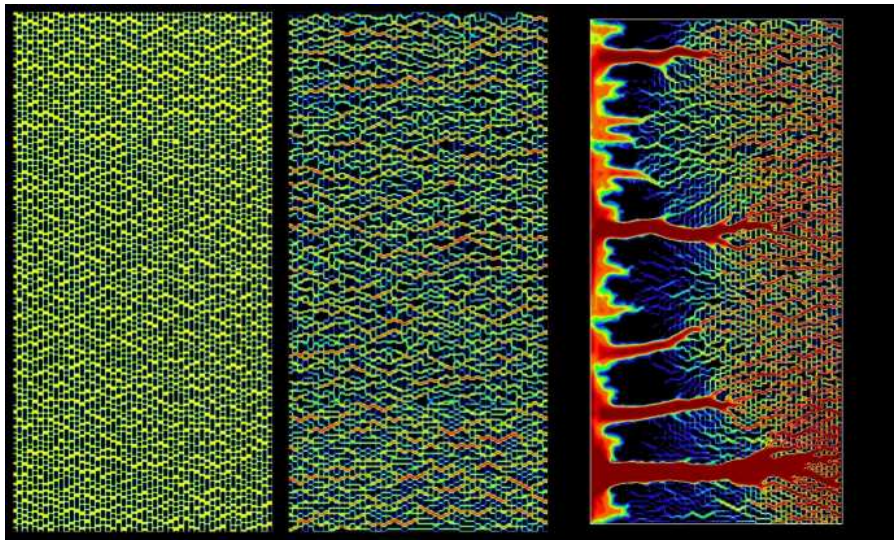}
 \caption{(Color online) Initial distribution of obstacles (dark pixels) in the artificial fracture (left); initial flow field (center);
and the flow field after an increase in mean aperture equal to its initial value, $\Delta \bar{h} = \bar{h}_0$ (right).
The flow field, $v^{2d}=\sqrt{{\bf v}^{2d} \cdot {\bf v}^{2d}}$, Eq.~\eqref{v2d}, is averaged over the local aperture.  The lightest/red shading indicates the highest velocity, while dark gray/blue indicates the lowest; the intermediate colors are yellow (higher) and green (lower). The flow direction is from left to right.}\label{inifull}
\end{figure}

A typical initial geometry ($\zeta=0.5$) is shown in the left panel of Fig.~\ref{inifull}; the integrated (two-dimensional) velocity field,
\begin{equation}\label{v2d}
{\bf v}^{2d}(x,y) = \int_{h_l(x,y)}^{h_u(x,y)}{\bf v}(x,y,z) dz,
\end{equation}
is shown in the center panel. Reactive fluid enters from left side and exits
from the right, while no-slip boundaries are imposed on the other surfaces.
The setup and initial topography resemble
the experiment in~\citet{Det03}, but here we allow both surfaces to dissolve, which speeds up dissolution and channel formation.
In this case the time-dependent erosion depth is defined in terms of the combined change in height of both upper and lower surfaces,
\begin{eqnarray}
\hspace{3em} \Delta h(x,y;t) &=& \left[h^u(x,y,t)-h^l(x,y,t)\right]  \\ &-& \left[h^u(x,y,0)-h^l(x,y,0)\right]. \nonumber
\end{eqnarray}
Initially there are no discernible channels (center panel) and the velocity field shows only short-range spatial correlations. During dissolution, large scale variations develop from small fluctuations in the initial porosity. In the final stages of dissolution (right panel in
Fig.~\ref{inifull}) the channeling is very distinct, but the size of these
spontaneously formed channels are not related to the initial pore size distribution, which is highly uniform.
The growth in mean aperture,
\begin{equation}
\Delta{\bar h}(t) = \frac{1}{L W} \int_0^L \int_0^W \Delta h(x,y,t) \mbox{d}y \mbox{d}x,
\end{equation}
will sometimes be used as a measure of elapsed time; it is then normalized by the initial mean aperture $\bar{h}_0$.

\section{Channeling as a function of P\'eclet and Damk\"ohler number}\label{sec:channels}

\begin{figure*}
\center\includegraphics[width=42pc]{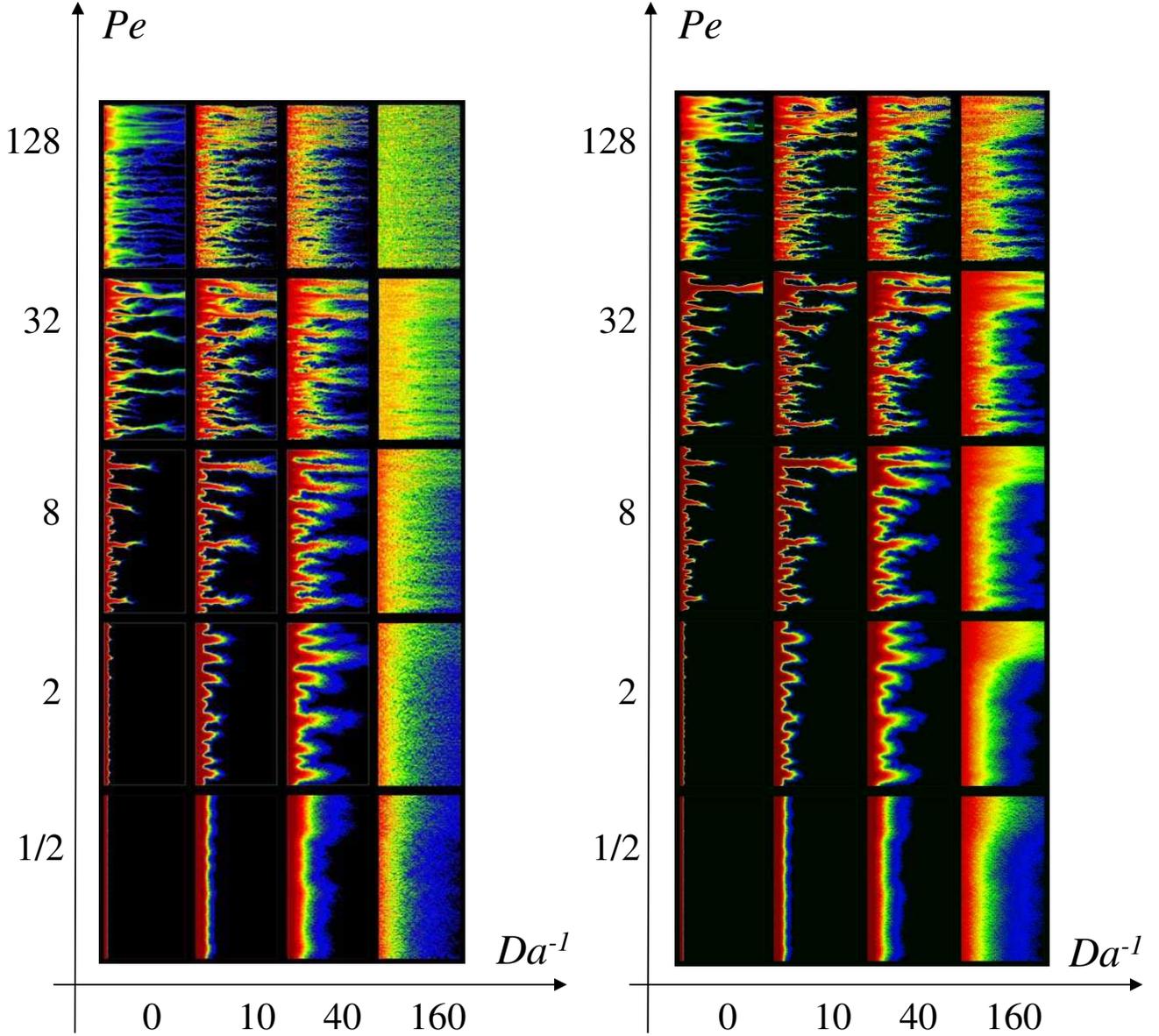}
\caption{(Color online) Erosion of the lower surface (initially flat) 
at constant pressure drop (left) and constant injection rate (right), for different
P\'eclet and Damk\"ohler numbers. The lightest/red shading indicates the deepest erosion while dark gray/blue indicates the least erosion; the intermediate colors are yellow (higher) and green (lower). The flow direction is from left to right.}\label{AllPen}
\end{figure*}

Figure~\ref{AllPen} illustrates typical dissolution patterns at a mean erosion depth $\Delta \bar{h} =2 \bar{h}_0$, over a range of different P\'eclet ($Pe$) and Damk\"ohler ($Da$) numbers. Changes in erosion patterns map more uniformly to variations in the inverse Damk\"ohler number, $Da^{-1}$, than variations in $Da$. We therefore use $Da^{-1}$ as the independent variable in most of our plots, although, in conformity with normal practice, we discuss the results in terms of variations in $Da$.

For small $Pe$ and large $Da$ the reactant saturates ($\rr=0$) near the injection face. After a fast initial dissolution of material at the fluid inlet, the reaction front propagates extremely slowly, as there is almost no unsaturated fluid penetrating inside the fracture. On the other hand, when the reaction rate is sufficiently slow ($Da < 1/100$), or the flow rate sufficiently high ($Pe > 500$), unsaturated fluid penetrates deep inside the fracture and the whole sample dissolves almost uniformly. Channeling is observed for moderate values of P\'eclet and Damk\"ohler numbers, $Pe \sim 10$ and $Da > 1/100$. 
Here nonlinear feedback plays a decisive role. A perturbation in the reaction rate at the dissolution front increases (for example) the local permeability, which in turn increases solute transport and therefore the local dissolution rate. The increasing flow rate reinforces the initial perturbation and the front becomes unstable, developing pronounced channels where the majority of the flow is focused, while most of the pore space is eventually bypassed. The interaction between channels, important in the later stages of dissolution, is discussed in Sec.~\ref{sec:worms}.

Above a threshold Damk\"ohler number, $Da > 1$, the erosion patterns are largely determined by the P\'eclet number. Additional data (not shown) demonstrates that there is no difference in dissolution patterns at $Da = 1$, $Da = 10$ and $Da \rightarrow \infty$, except at small P\'eclet numbers, $Pe < 1$. A similar range of Damk\"ohler number ($0.1 < Da < \infty$), has been reported in other numerical studies~\citep{Ste90} as a regime where the length of a single dissolving channel in a two-dimensional porous medium becomes independent of reaction rate; we will return to this point in Sec~\ref{sec:Deff}. In the mass-transfer-limited case, where the surface reaction rate is high enough that the overall dissolution process no longer depends on reaction rate, the stochastic modeling may be simplified by imposing an absorbing boundary condition, $\rr=0$, at the fracture surfaces, corresponding to the limiting case $Da \rightarrow \infty$; these results are shown in the leftmost columns of Fig.~\ref{AllPen}. 

\begin{figure}[t]
\center\includegraphics[width=18pc]{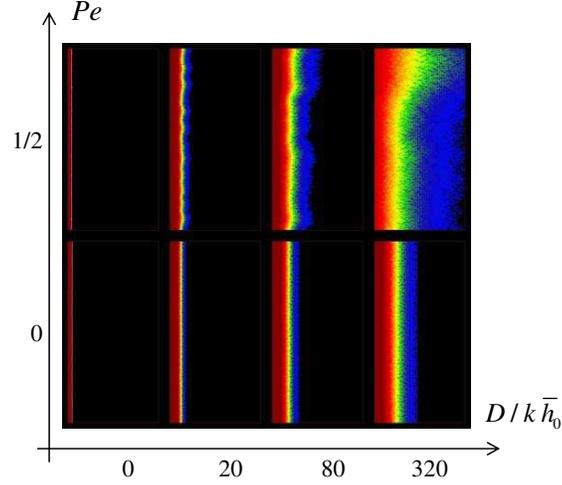}
\caption{(Color online) Dissolution patterns in the diffusive regime, for $Pe=1/2$ (upper row) and 
$Pe=0$ (lower row). The values of $(PeDa)^{-1} = D / k \bar{h}_0$ are marked.} \label{diffus}
\end{figure}

For smaller Damk\"ohler numbers, $Da<1$, the dissolution patterns become dependent on both $Pe$ and $Da$.
The P\'eclet number controls the number of channels, with the spacing between them decreasing with increasing
$Pe$. On the other hand, for fixed $Pe$, a decrease in Damk\"ohler number results in a more diffuse boundary between the channels and the surrounding porous matrix.
When the P\'eclet number is less than one, diffusive transport becomes more important than 
convection, even in the flow direction. In this regime, the dissolution patterns are determined by the
product of P\'eclet and Damk\"ohler number,
\begin{equation}
PeDa = \frac{k \bar{h}}{D},
\end{equation}
which gives the relative magnitude of reactive and diffusive fluxes. Figure~\ref{diffus} compares the dissolution 
patterns for $Pe = 1/2$ with those for the purely diffusive case, $Pe = 0$. The differences are rather slight, except at very small reaction rates; the reaction front propagates  slowly and stably inside the fracture, with the penetration length decreasing with increasing $PeDa$.

\begin{figure}[t]
\center\includegraphics[width=18pc]{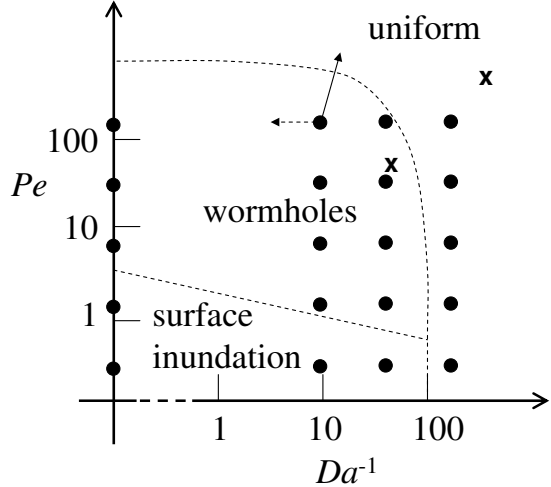}
\caption{Phase diagram describing characteristic dissolution patterns as a function
of P\'eclet and Damk\"ohler number. The points mark the values of P\'eclet and Damk\"ohler
number corresponding to the patterns shown in Fig.~\ref{AllPen}, and the crosses mark the KDP fracture experiments shown in Fig.~\ref{tails2}. Since the scales are logarithmic, the points corresponding to $Da^{-1} = 0$ are marked at $Da^{-1} = 0.1$, to which they are similar. The arrows indicate the direction the points move in the phase diagram during dissolution at constant pressure drop (solid arrow) and constant flow rate (dashed arrow).}\label{phasediag}
\end{figure}

These results are summarized in a phase diagram, Fig.~\ref{phasediag}, where
the values of P\'eclet and Damk\"ohler number corresponding to the patterns shown in
Fig.~\ref{AllPen} are marked, together with the points corresponding to the KDP fracture experiments by~\cite{Det03} (see Sec.~\ref{kdp}). Although the geometry of the
KDP fracture is different from the obstacle fracture geometries considered here, the
dissolution patterns captured at comparable $Pe$ and $Da$ are nevertheless similar. For example, the KDP dissolution pattern at $Pe=54$, $Da=0.018$ (Fig.~\ref{tails2}) may be compared with the artificial fracture at $Pe=32, Da=0.025$ (Fig.~\ref{AllPen}). 

A feature of reactive flows, as compared with other pattern forming systems, is that the dimensionless numbers characterizing the flow and reaction rates are changing throughout the course of the dissolution~\citep{Dac93b}. Indeed, in constant pressure drop simulations (left panel of Fig.~\ref{AllPen}), both the total flow and the mean aperture change during the course of dissolution; thus the point in the phase diagram representing the initial system moves towards larger $Pe$ and smaller $Da$ values as the dissolution progresses (solid arrow in Fig.~\ref{phasediag}).

Constant pressure drop conditions are representative of many groundwater flow systems, including the early stages 
of karstification~\citep{Drey90}. However, 
in a number of technological applications, {\em e.g.} acidization of petroleum reservoirs~\citep{Eco2000}, the control variable is the injection rate of reactive fluid, $Q$. In that case, the P\'eclet number remains constant throughout the dissolution process, since
\begin{equation}
Pe = \frac{\bar{v} \bar{h}}{D} = \frac{Q}{W D}.
\end{equation}
On the other hand, the Damk\"ohler number then increases in proportion to $\bar{h}$ 
(dashed arrow in Fig.~\ref{phasediag}), since
\begin{equation}
Da = \frac{k}{\bar{v}} = \frac{k \bar{h} W}{Q} .
\end{equation}
The results of the constant injection rate simulations are presented in the right panel of Fig.~\ref{AllPen}. While 
the differences are not dramatic, channels formed in constant pressure drop simulations
are noticeably more diffuse at their tips. This is consistent with the analysis in Sec.~\ref{front}, where it is shown that 
the thickness of the dissolution front near the channel tip is proportional to $Da^{-1}$. The Damk\"ohler number decreases in the course of constant pressure drop simulations while it increases during constant flow rate runs, which leads to different front thicknesses at the tip. 
This observation agrees with laboratory experiments~\citep{Hoe88}, where it was observed that
wormholes in acidized limestone formed by constant pressure drop dissolution become more highly branched at later times than those formed at constant flow rate. 

\begin{figure}[t]
\center\includegraphics[width=18pc]{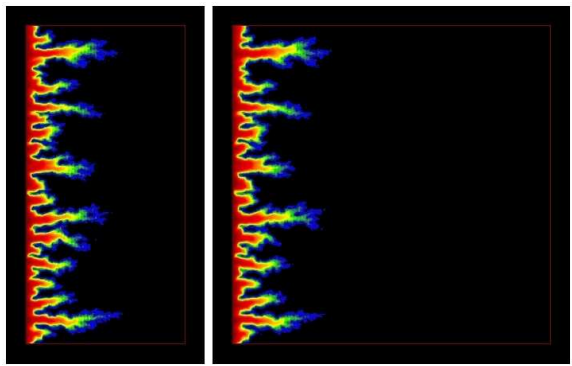}
\caption{(Color online) Comparison of dissolution patterns at $Pe=8$, $Da=1/10$ between the standard domain $200 \times 400$ pixels (left), used in Fig.~\ref{AllPen}, and a larger, $400 \times 400$ domain with the same topography (right).} \label{largesmall}
\end{figure}

We have verified that the results shown in Fig.~\ref{AllPen} are independent of the length of the fracture domain. In  Fig.~\ref{largesmall}, the dissolution pattern in a $200 \times 400$ pixel fracture are compared with a longer $400 \times 400$ domain at the same P\'eclet and Damk\"ohler numbers, $Pe=8$, $Da=1/10$. The initial topographies of both systems were identical in the $200 \times 400$ pixel inlet region, and the comparison was made when the same volume of material had been eroded from each sample. We find that the  dissolution patterns in both systems  are very similar, up to the point where the dominant channels reach the outflow of the smaller domain, as can be seen in Fig.~\ref{largesmall}.  In Sec.~\ref{sec:worms}, we describe how the simple hierarchical growth pattern of the competing channel system then becomes disrupted,  with large pressure gradients developing at the tips of the leading channels, causing them to split (see also Fig.~\ref{dissol}). Here we simply wish to point out that, in the wormholing regime, the results shown in Fig.~\ref{AllPen} are insensitive to further increases in the length of the domain. However, in the uniform-dissolution regime, the overall length of the fracture is important. The average undersaturation decays exponentially along the flow direction and so will eventually reach a region where the solution is saturated and no dissolution occurs. In this sense it seems that the distinction between uniform dissolution and surface inundation is merely a matter of scale. When the length of the fracture is comparable to the depth of penetration, dissolution appears uniform, but if the fracture is much longer, then dissolution is limited to a small region (relative to the length) near the inlet.

\section{Effective reaction rate}\label{sec:Deff}

The phase diagram in Fig.~\ref{phasediag} can be better interpreted in terms of an effective reaction rate, which takes account of the interplay between mass transfer and chemical kinetics.
In a two-dimensional description of dissolution~\citep{Detwiler2007}, the fracture is locally approximated by two parallel plates separated by a distance $h(x,y,t)$. In the reaction-limited regime, diffusive timescales are much shorter than reactive ones, ${k h}/{D} \ll 1$, and the concentration field is almost uniform across the aperture. The reactant flux is then given by
\begin{equation}
J=k C_0 \approx k {\bar C}
\label{j1}
\end{equation}
where ${\bar C}(x,y)$ is the cup-averaged concentration.
In the opposite limit, ${k h}/{D} \gg 1$, dissolution becomes mass-transfer limited and an absorbing boundary condition, $C_0=0$, may be assumed at the fracture walls. The dissolution flux,
\begin{equation}
J= k_t {\bar C},
\label{j2}
\end{equation}
is determined by the mass transfer coefficient $k_t$, 
\begin{equation}
k_t=Sh \frac{D}{d_h},
\label{kt}
\end{equation}
where $d_h$ is the hydraulic diameter of the system and $Sh$ is the Sherwood number. For parallel plates $d_h=2 h$ and $Sh=7.54$~\citep{Bir60}.

In the general case when ${k h}/{D} \sim 1$, the reactant flux may expressed 
in terms of ${\bar C}$ by 
equating the dissolution flux \eqref{j1} to the mass-transfer flux \eqref{j2}
\begin{equation}
J = k C_0 = k_t ({\bar C}-C_0)
\end{equation}
Solving for $C_0$ in terms of ${\bar C}$ gives
\begin{equation}
J = k_{eff} {\bar C}
\end{equation}
with
\begin{equation}
k_{eff}  = \frac{k k_t}{k+k_t}
\label{keff}
\end{equation}

There are two approximations made here. First,  the Sherwood number \eqref{kt} depends, in general, on the reaction rate, $k$. However, the variation in $Sh$ is relatively small~\citep{Hayes1994,Gupta2001}, bounded by two asymptotic limits:
constant flux, where $Sh=8.24$ for parallel plates, and constant concentration, where $Sh=7.54$. Here we take the approximate value $Sh=8$ in order to estimate $Da_{eff}$.
Second, we have neglected entrance effects, which otherwise make the Sherwood number dependent on the
distance from the inlet, $x$. However, the entrance length (defined as the distance 
at which the Sherwood number attains a value within $5 \%$ of the asymptotic one)
is negligibly small, $L_x \approx 0.008 d_h Pe$~\citep{Ebadian1998}, at least in the early stages of the dissolution.
Expressions for the effective reaction rate coefficient analogous to \eqref{keff}
have been proposed previously~\citep{Rickard1983,Dre96,Panga2005,Detwiler2007}, but a
somewhat different approach was employed by~\citet{Gro94b} and~\citet{Han98}
where the smaller of the two rates $k$ and $k_t$ was used for $k_{eff}$.

\begin{figure}
\center\includegraphics[width=18pc]{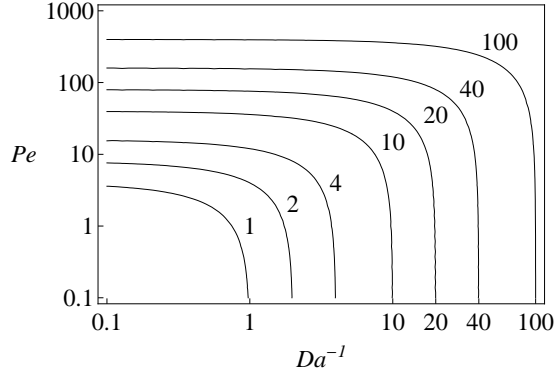}
\caption{Contours of the inverse effective Damk\"ohler number, $Da_{eff}^{-1}$, as a function of $Da^{-1}$ and $Pe$.}\label{pdaeff}
\end{figure}

\begin{figure}[t]
\center\includegraphics[width=18pc]{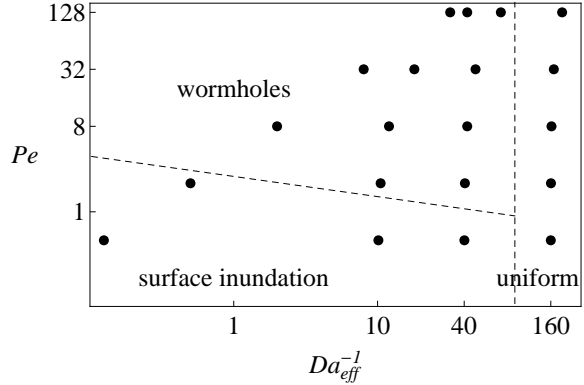}
\caption{Phase diagram describing characteristic dissolution patterns as a function
of P\'eclet and effective Damk\"ohler number.}\label{diageff}
\end{figure}

The effective reaction rate, $k_{eff}$ (Eq.~\ref{keff}, can be used to construct an effective
D\"amkohler number, $Da_{eff}={k_{eff}}/{\bar{v}}$,
\begin{equation}
Da^{-1}_{eff} = Da^{-1}+ \frac{2 Pe}{Sh},
\label{daeffinv}
\end{equation}
which remains finite even when the reaction rate becomes very large, $Da \rightarrow \infty$; contours of constant $Da_{eff}$ are shown in Fig.~\ref{pdaeff}. The phase diagram of dissolution patterns in the $Pe-Da_{eff}$ plane, shown in Fig.~\ref{diageff}, is simpler than the corresponding phase diagram in $Pe-Da$ (Fig.~\ref{phasediag}). In particular, uniform dissolution can now be uniquely associated
with small $Da_{eff}$, whereas in $Pe-Da$ variables it corresponds to either small $Da$ or large $Pe$. The introduction of $Da_{eff}$ also explains the independence of the dissolution patterns on the microscopic Damk\"ohler number in the $Pe > 1, Da > 1$ regime discussed in Sec.~\ref{sec:channels}. At higher P\'eclet numbers a large change in $Da$ corresponds to 
a relatively small change in $Da_{eff}$. For example, at $Pe=32$, $Da_{eff}$ changes from $0.125$ when $Da\rightarrow\infty$ to $0.111$ when $Da = 1$.

The general features of the phase diagram agree with experimental and numerical studies of wormhole formation in quasi-two-dimensional porous media~\citep{Golfier2002}. However, in these simulations the transitions between different dissolution patterns corresponded to fixed values of either P\'eclet
or Damk\"ohler number; in other words, the boundaries in the $Pe-Da_{eff}$ phase diagram were perpendicular to the axes. In our case the line between surface inundation and the wormhole regime is not horizontal; at low P\'eclet numbers, the critical value of P\'eclet number at which channels are  spontaneously formed decreases with decreasing $Da_{eff}$. The discrepancies at low $Pe$ may be caused by a transition to a more three-dimensional flow in our fracture simulations. Although the initial geometry shares many features with the two-dimensional porous media considered in other studies~\citep{Golfier2002,Panga2005}, the
third dimension plays a more significant role as the dissolution proceeds, particularly
in a large $Da_{eff}$, small $Pe$ regime, where the penetration of the reactive fluid is very small.
In this regime, the solutional widening of the fracture at the inlet can be more than one order of magnitude larger than the mean aperture growth in the system and thus the system ceases to be quasi two dimensional. 

\section{Front thickness}\label{front}

A detailed examination of Fig.~\ref{AllPen} reveals a number of qualitative features of the developing channels. At high Damk\"ohler numbers and low P\'eclet  numbers the channels are very distinct, with sharp well-formed boundaries between dissolved and undissolved material. As the P\'eclet number increases the channels become more diffuse, but only in the flow direction. The lateral thickness of the channels is almost unchanged, while along the flow direction a sharp transition is replaced by a gradual change in dissolution depth, which takes place over almost the whole channel length; this is especially pronounced above $Pe = 30$ in the constant pressure-drop case and $Pe=100$ in the constant flow rate case. The second qualitative feature is that at smaller Damk\"ohler numbers,  the channels become laterally diffuse as well, as can be seen from the broad blue regions at the dissolution front when $Da < 0.1$. Insight into the characteristics of channel formation can be gained from a simple model based on a Darcy-scale description of a wormhole. 

\begin{figure}
\center\includegraphics[width=18pc]{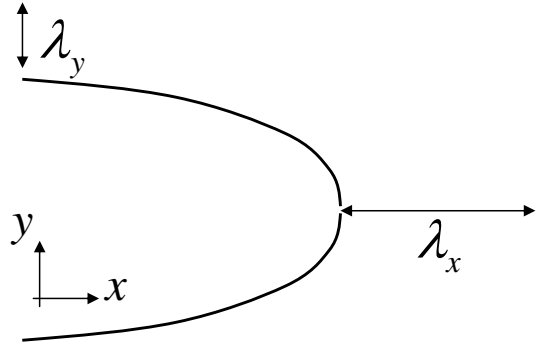}
\caption{Sketch indicating the characteristic dimensions of a wormhole front. The parameters $\lambda_x$ and $\lambda_y$ indicate the extent of the dissolution into the porous matrix.}\label{transport}
\end{figure}

Consider the tip of a channel containing reactive fluid, with depth-averaged concentration ${\bar \rr}$,
entering the undissolved medium. At the leading edge of the wormhole, the Darcy-scale equation for reactant transport takes the form~\citep{Lichtner1988,Ste90},
\begin{equation}
\frac{\partial {\bar \rr}}{\partial t} = D_x \frac{\partial^2 {\bar \rr}}{\partial x^2}  - v \frac{\partial {\bar \rr}}{\partial x} - \frac{2 k_{eff}}{\bar{h}_0} {\bar \rr},
\label{effective}
\end{equation}
where $D_x = D(1 + \beta_x Pe)$ is the dispersion coefficient along the flow direction, which includes the effects of fluctuations in the fluid velocity through the coefficient $\beta_x \sim 0.5$ \citep{Panga2005}. The last term describes the concentration loss due to dissolution in the
medium ahead of the reaction front, assuming the fracture here is undissolved
and may be approximated by two parallel surfaces separated by $\bar{h}_0$ (Sec.~\ref{sec:Deff}).
Then the erosion flux at both upper and lower surfaces is $k_{eff} {\bar \rr}$ and the rate of change in concentration is $-2k_{eff} {\bar \rr}/\bar{h}_0$. 
The stationary solution of \eqref{effective} is~\citep{Ste90},
\begin{equation}
{{\bar \rr}}(x)={{\bar \rr}}_{tip} e^{- x/\lambda_x}
\end{equation}
where ${\bar \rr}_{tip}$ is the reactant concentration at the tip of the channel.

The characteristic thickness of the dissolution front, $\lambda_x$, is given by~\citep{Lichtner1988,Ste90}
\begin{equation}
\lambda_x = \frac{2D_x}{v} \left[ \left(1+ \frac{8 D_x
k_{eff}}{\bar{h}_0 v^2}\right)^{1/2}-1 \right]^{-1}.
\label{thickness}
\end{equation}
Along the flow direction, convective effects are typically much stronger than diffusive ones, and Eq.~\eqref{thickness} simplifies,
\begin{equation}\label{thickx}
\lambda_x = \frac{\bar{h}_0}{2 Da_{eff}} = \frac{\bar{h}_0}{2} \left(\frac{1}{Da}+\frac{2 Pe}{Sh}\right),
\end{equation}
where $\lambda_x$ is the thickness of the front ahead of the wormhole tip  (see Fig.~\ref{transport}).
Convective effects are small in directions transverse to the flow, and setting $v = 0$ in the analogue of Eq.~\eqref{effective}, gives the transverse thickness of the reaction front,
\begin{equation}\label{thicky}
\lambda_y =  \left(\frac{D_y{\bar h}_0}{2 k_{eff}}\right)^{1/2} =  {\bar h}_0 \left( \frac{1 + \beta_y Pe}{2 Pe Da_{eff}} \right)^{1/2},
\end{equation}
where the dispersion coefficient $\beta_y \sim 0.1$ \citep{Panga2005}.

In the reaction-limited regime ($PeDa \ll 1$), $\lambda_x \approx {\bar h}_0/2Da$, independent of P\'eclet number. This scaling prediction can be observed in Fig.~\ref{AllPen}, most clearly at $Da = 1/40$, for both constant pressure drop and constant flow rate conditions. The extent of the front roughly corresponds to the blue (color figure) or dark grey regions in front of the wormhole. At $Da = 1/40$ these diffuse regions extend roughly 20--30\% of the length of the channel ($\sim 40-60 {\bar h}_0$), independent of $Pe$. The scaling with $Da$ is approximately linear, which can be seen most clearly when comparing $Da = 1/40$ with $Da = 1/10$ at constant pressure drop; at constant flow rate the extent of the dissolution front is too small to measure at $Da = 1/10$. In the mass-transfer-limited regime ($PeDa \gg 1$), $\lambda_x \approx {\bar h}_0 Pe/Sh$. The extent of the reaction front at $Da = \infty$ (Fig.~\ref{AllPen}) depends roughly linearly on $Pe$, in agreement with the simple scaling law. A quantitative comparison is suspect for several reasons: the difficulty in defining and measuring the extent of the front, changes in P\'eclet and Damk\"ohler numbers as dissolution proceeds, and local variations in fluid velocity. Nevertheless, the Darcy-scale model qualitatively explains the key results of the simulations shown in Fig~\ref{AllPen}.

The behavior of the transverse thickness of the front is more complicated, because of the dispersion term in Eq.~\eqref{thicky}. In the reaction-limited regime, $\lambda_y =  {\bar h}_0 \left[(1 + \beta_y Pe)/(2 Pe Da)\right]^{1/2}$, and the line $PeDa = 1$ is the division between sharply defined and laterally extended wormholes (see Fig.~\ref{AllPen}). At sufficiently high P\'eclet numbers the dispersion term should eliminate the dependence of $\lambda_y$ on $Pe$, but the value of $PeDa$ is then too large for an observable lateral extension of the reaction front. An exception may be the case $Pe = 128$, $Da = 1/160$ with constant flow-rate conditions (right panel of Fig~\ref{AllPen}). This is a transition case between wormholing and uniform erosion, and here we can see a significant lateral spreading of the front. Much more clearly defined is the growth in lateral thickness as $PeDa$ gets smaller. This again can be most readily seen at $Da = 1/40$; here the lateral extension of the front gets more pronounced as $Pe$ is reduced. Finally, in the mass-transfer-limited case, $\lambda_y = {\bar h}_0 \left[(1+\beta_y Pe)/Sh\right]^{1/2} \lesssim {\bar h}_0$, and the reaction front is always limited to the region ahead of the tip, as is observed at $Da = \infty$ (Fig.~\ref{AllPen}).

\citet{Panga2005} argue that the ratio of these two length scales
\begin{equation}
\Lambda = \frac{\lambda_y}{\lambda_x}\label{eq:Lambda}
\end{equation}
determines the aspect ratio of the wormhole, with the strongest wormholing predicted to occur when $\Lambda \approx 1$. The latter criteria matches quite well with the patterns observed in the simulations. However, examination of Fig.~\ref{AllPen} shows that the aspect ratio of the wormholes does not correlate well with $\Lambda$. For example, at constant $Pe$, the wormhole diameter is only weakly dependent on $Da$ (Fig.~\ref{AllPen}), and the aspect ratio remains more or less constant, whereas the front thickness increases considerably with decreasing $Da$. Thus the aspect ratio of the reaction front, as measured by $\Lambda$, does not necessarily control the aspect ratio of the wormhole itself. The dependence of the wormhole shape on P\'eclet number is analyzed from a different perspective in the next section, based on a microscopic mass balance within the wormhole.

\section{Channel shape}\label{shape}

\begin{figure}
\center\includegraphics[width=18pc]{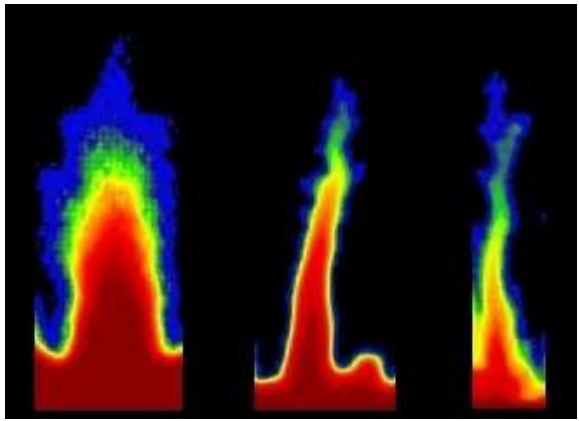}
\caption{(Color online) Examples of channels of a similar length at different P\'eclet numbers (from left to right) $Pe=2$, $Pe=8$, and $Pe=32$. At the two higher P\'eclet numbers we used $Da = \infty$, but at $Pe = 2$ we used a lower Damk\"ohler number, $Da = 0.1$, since the channels do not form at $Da = \infty$ in this case.
}\label{diameters}
\end{figure}

\begin{figure}
\center\includegraphics[width=20pc]{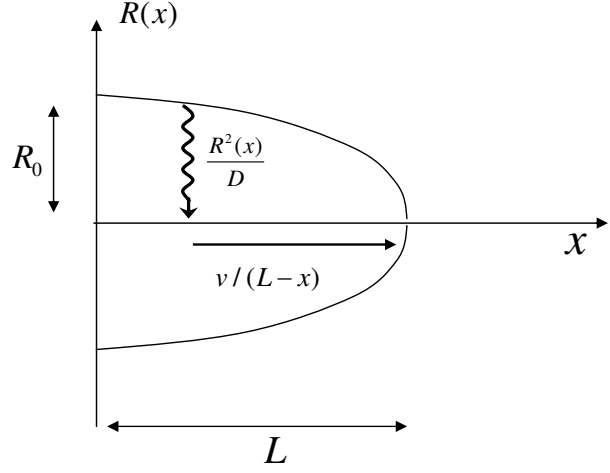}
\caption{The radius of a wormhole, $R$, as a function of the longitudinal coordinate, $x$. The arrows mark
the characteristic time scales of diffusive and convective transport.
}\label{wormhole}
\end{figure}

The aspect ratio of the dissolving channels increases with increasing P\'eclet  number as can be seen in Fig.~\ref{diameters}. For a fixed length, the typical wormhole diameter decreases approximately as $Pe^{-1/2}$. This scaling can be understood by noting \citep{Ste90}, that, in the mass-transfer-limited regime, the shape of the channel depends on the interplay between diffusive transport normal to the flow, and convective transport along the flow direction.
The channel boundary is parameterized by the curve $R(x)$, where $R$ is the distance of a point on the boundary from the channel center line; the geometry is illustrated in Fig.~\ref{wormhole}. The depleted concentration at the point $\{x,R(x)\}$ diffuses towards the center of the channel with a timescale that can be estimated by solving a two-dimensional diffusion equation for the concentration $C(r,t)$, inside a circle with an outer boundary condition $C(R,t) = 0$. This gives  an asymptotic (in time) dependence of the concentration at the center line
\begin{equation}
C(0,t) \sim e^{-a_0^2 Dt/R^2},
\end{equation}
where $a_0 \approx 2.4$ is the first zero of the Bessel function $J_0$. Thus the characteristic time for depleted reactant to diffuse from the channel boundary $R(x)$, is given by
\begin{equation}
\Delta t = \frac{R^2(x)} {a_0^2 D}.
\end{equation} 
The diffusing concentration field is
simultaneously advected by $v \Delta t$ to the tip of the wormhole located at $\{L,0\}$ (Fig.~\ref{wormhole}). Equating the two timescales,
\begin{equation}
\frac{L-x}{v} = \Delta t =  \frac{R(x)^2}{a_0^2 D},
\end{equation}
leads to an expression for the channel shape
\begin{equation}
R(x)=\left(\frac{a_0^2 D(L-x)}{v}\right)^{1/2}  = R_0 \sqrt{1-x/L},
\end{equation}
where
\begin{equation}\label{eq:R_0}
R_0= \left(\frac{a_0^2 \bar{h}_0 L}{Pe}\right)^{1/2}
\end{equation}
is the radius of the wormhole (Fig.~\ref{wormhole}).

The analysis in this section uses a microscopic model of reactant transport {\em within} the wormhole, in contrast to the previous section (Sec.~\ref{shape}) where we considered reactant transport in the porous matrix. Thus, in the flow direction we assume that convection is dominant, whereas in the transverse directions the transport is molecular diffusion rather than dispersion. This is based on a picture of a wormhole as a region of low (or vanishing) porosity, so that the flow is roughly parabolic, constrained by the upper and lower fracture surfaces and by the boundary of the wormhole; the fluid velocity in the wormhole is much higher than in the surrounding matrix. Along the flow direction, diffusion is enhanced by Taylor dispersion,
\begin{equation}
D_\parallel = D(1 + \beta_T Pe^2),
\label{eq:TaylorD} 
\end{equation} 
where $\beta_T$ is a coefficient that depends on geometry, but is $\sim 0.005$. At moderate P\'eclet numbers, $Pe > 10$, Taylor dispersion is comparable to or larger than molecular diffusion, but even at the highest P\'eclet numbers in these studies ($Pe \sim 100$) dispersion makes a negligible contribution to the axial transport of reactant. The timescale for convective transport over the wormhole length $L$ is still much smaller than the diffusive timescale, even if Taylor dispersion is included,
\begin{equation}
\frac{t_{diff}}{t_{conv}} = \frac{{\bar v} L}{D_\parallel} \approx \frac{L}{{\bar h}_0 \beta_T Pe} \gg 1.
\end{equation} 
The last inequality is valid up to at least $Pe \sim 10^2$ for channel lengths in excess of $10 {\bar h}_0$.

\begin{figure}
\center\includegraphics[width=18pc]{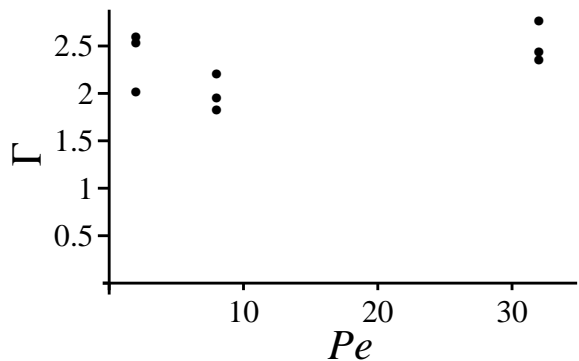}
\caption{The dimensionless parameter,  $\Gamma = R_0 \left( Pe / \bar{h}_0 L \right)^{1/2}$, characterizing the wormhole shape, as a function of a P\'eclet number, Eq.~\eqref{eq:G}.
}\label{width}
\end{figure}

In this simple wormhole model, Eq.~\eqref{eq:R_0}, the radius of the channel scales like $Pe^{-1/2}$, while its shape is parabolic. To connect the theory with the numerical simulations, we have calculated the dimensionless quantity,
\begin{equation}\label{eq:G}
\Gamma = R_0\left(Pe/\bar{h}_0 L\right)^{1/2},
\end{equation} 
which Eq.~\eqref{eq:R_0} predicts will have a universal value of $a_0 \approx 2.4$. In making these comparisons we must take into account that the analysis leading to Eq.~\eqref{eq:R_0} considers only individual channels. However, channel competition is an essential component of the overall dynamics~\citep{Szy06}, in which the longer channels drain flow from the shorter ones, limiting their growth. A detailed analysis of channel competition is given in Sec.~\ref{sec:worms}; here we aim to limit the effects of channel competition by focusing on the longest channels, which remain active throughout the dissolution process.
Fig.~\ref{width} shows the dimensionless ratio $\Gamma$, Eq.~\eqref{eq:G}, for dissolving fractures in the mass transfer limit. In accordance with the above remarks, only the three longest channels in each fracture were measured. The numerical results confirm that $\Gamma$ is nearly universal, independent of P\'eclet number and the choice of channel. Moreover, the numerical values of $\Gamma$ are close to 2.4, but this may be accidental given the limited precision of the numerical data and the simplicity of the model.

In the opposite case, when the process is reaction-limited and $Pe \gg 1$, the reactant concentration in the entire wormhole is nearly uniform and equal to the inlet concentration, $C_{in}$. Interestingly, in this limit, the shape of the wormhole is also parabolic~\citep{Nilson1990}.

\section{Permeability evolution and optimal injection rates}\label{sec:perm}

\begin{figure*}
\center\includegraphics[width=36pc]{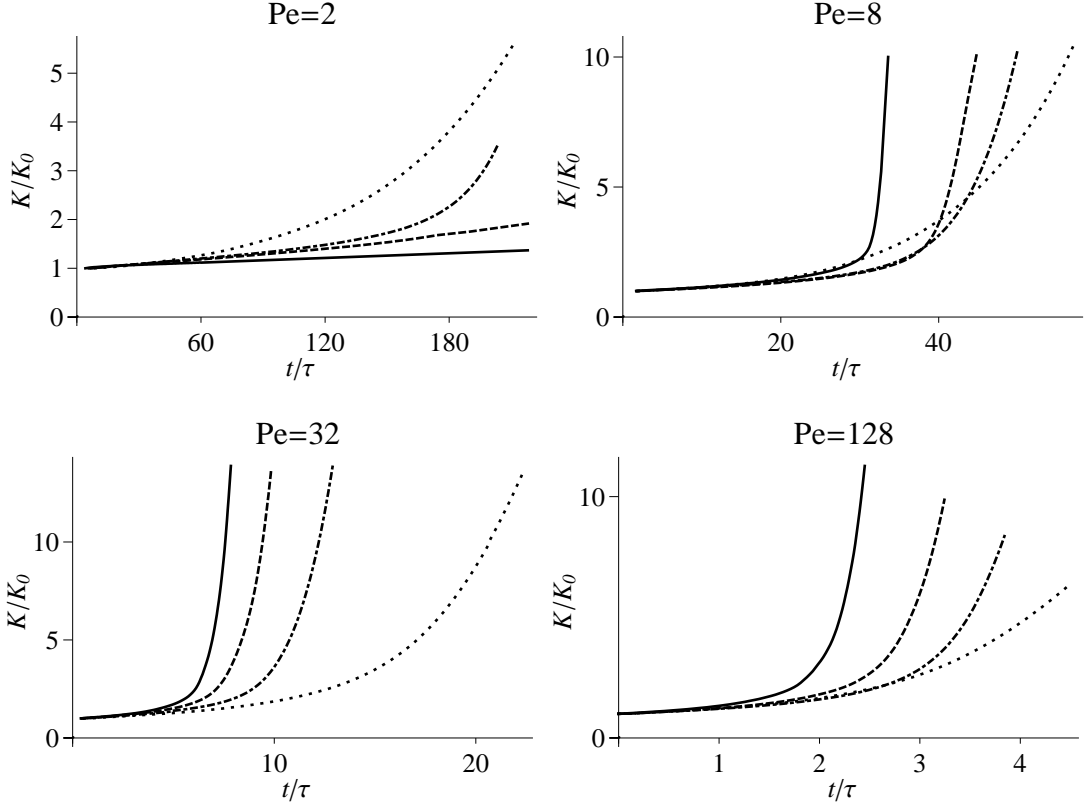}
\caption{Permeability as a function of time during dissolution
at constant pressure drop; results are shown for $Pe=2$, $8$, $32$, and $128$. The lines correspond to:
$Da = 1/160$ (dotted), $Da = 1/40$ (dash-and-dotted), $Da = 1/10$ (dashed), and
$Da =\infty$ (solid).
}\label{flowst}
\end{figure*}

The evolving permeability in a dissolving fracture can be defined by the relation
\begin{equation}
Q = -\frac{K W \bar{h}_0 \nabla p}{\mu}
\end{equation}
where $K$ is the permeability, $\mu$ is the viscosity and $W \bar{h}_0$ is the initial
cross-sectional area of the fracture. Figure~\ref{flowst} shows the permeability as a function of time for dissolution at constant pressure drop.
At low P\'eclet numbers ($Pe=2$ in Fig.~\ref{flowst}), the flow rate through
the sample only increases significantly at very small Damk\"ohler numbers, where the dissolution is uniform throughout the fracture.
At higher Damk\"ohler numbers, surface inundation occurs and the flow
rate remains nearly constant throughout the simulation.
The behavior changes dramatically as the P\'eclet number is increased. At $Pe>20$, the
dependence of the flow rate on Damk\"ohler number is reversed: now the flow
increases most rapidly for larger values of $Da$. This is because wormhole
formation, triggered in this parameter range, becomes amplified
as the reaction rate increases.

A characteristic feature of systems that exhibit channeling is the rapid growth of the flow rate when the
dominant channels break through to the outflow end of the fracture; breakthrough is indicated by the
near vertical lines in Fig.~\ref{flowst}. At moderate P\'eclet numbers, $Pe \approx 10$,
uniform dissolution and wormholing compete with each other, as can be seen from the permeability evolution at P\'eclet number $Pe = 8$ (Fig.~\ref{flowst}). The permeability at first increases fastest at the lowest Damk\"ohler number $Da=1/160$, since the unsaturated fluid penetrates deeper inside the sample. However, once channels begin to form, the permeability in the higher Damk\"ohler systems increases very rapidly, and eventually overtakes the $Da=1/160$ system.

\begin{figure}
\center\includegraphics[width=36pc]{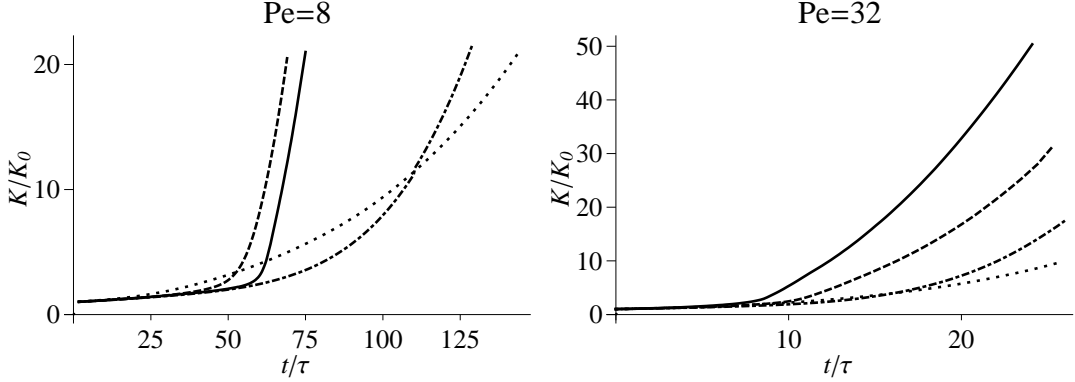}
\caption{Permeability as a function of time during dissolution
at constant flow rate; results are shown for $Pe=8$ and  $32$. The lines correspond to:
$Da = 1/160$ (dotted), $Da = 1/40$ (dash-and-dotted), $Da = 1/10$ (dashed), and
$Da =\infty$ (solid).
}\label{flowsq}
\end{figure}

At constant flow rate, shown in Fig.~\ref{flowsq}, the permeability increases more slowly overall than at constant pressure drop, where dissolutional opening of the fracture is enhanced by the increasing flow rate (note the different time scales in Figs.~\ref{flowst} and ~\ref{flowsq}). Positive feedback is particularly strong near breakthrough, which  manifests itself in steeper $K(t)$ curves in the constant pressure drop simulations. Another difference between the constant pressure drop and constant flow rate can be observed at $Pe=8$. At constant pressure drop the permeability growth is faster for $Da\rightarrow\infty$
than for $Da=0.1$, whereas for constant flow rate this order is reversed. This is again connected with the increasing flow rate in the course of  dissolution at constant pressure drop.  Since $Pe=8$, $Da\rightarrow\infty$ lies near the borderline between wormholing and surface inundation regimes, even small differences in flow have a pronounced impact on the speed at which channels propagate.

\begin{figure*}
\center\includegraphics[width=38pc]{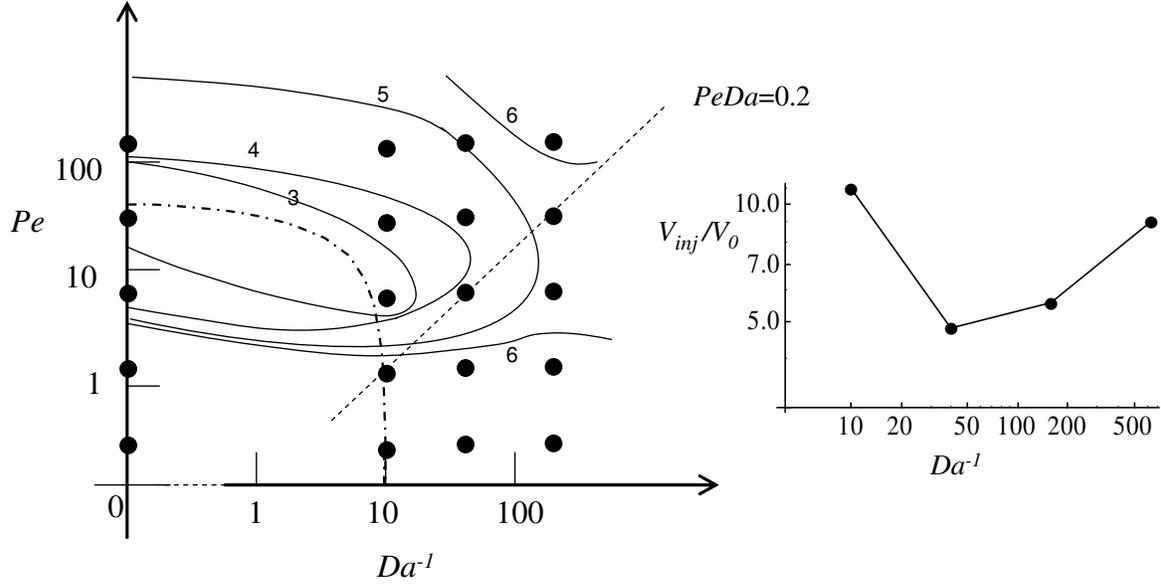}
\caption{Contour plot of the volume of reactant $V_{inj}$ needed for a 20-fold increase in permeability. The contours are normalized by $V_0$, the volume of reactant needed to dissolve a solid volume equal to the initial pore space in the fracture. The dashed line corresponds to varying the injection rate at $PeDa = 0.2$, and the inset shows the cross-section of the $Pe-Da^{-1}$ surface along this line. The dot-dashed line corresponds to $Da_{eff}=1/10$ and indicates a near optimum injection condition in the mass-transfer-limited regime.
}\label{porevolumes}
\end{figure*}

A quantitative description of channeling under constant 
flow conditions is also important in carbonate reservoir stimulation, where the relevant question is how to get the maximum increase of permeability for a given amount of reactive fluid. Numerical and experimental investigations of reactive flows in porous
media~\citep{Fredd1998,Golfier2002,Panga2005,Kalia2007,Cohen2008} suggest that there exists an optimum injection rate, which
maximizes the permeability gain for a given amount of fluid. If the injection rate is relatively small, surface inundation occurs 
and the increase in permeability is small. On the other hand, for very large injection rates, the reactant is exhausted on a uniform opening of the fracture, which is inefficient in terms of permeability increase. The optimum flow rate must give rise to spontaneous channeling, since
the reactant is then used to create a small number of highly permeable channels, which transport the flow most efficiently. 
To quantify the optimization with respect to $Pe$ and $Da$, we measured the total volume of reactive fluid, $V_{inj}$, that must be injected into the fracture in order to increase the overall permeability, $K$, by a factor of $20$. In the constant injection rate case,
\begin{equation}
V_{inj} = Q T \tau = Pe W D T  \tau
\end{equation}
where $T$ is the time needed for the given permeability increase, measured in units of $\tau$. Inserting the definition of $\tau$ \eqref{tau} gives 
\begin{equation}
V_{inj} = T Pe \frac{\bar{h}_0}{L} V_{0},
\end{equation}
where
\begin{equation}
V_{0} = W L \bar{h}_0 \frac{c_{sol}}{{\rr}_{in}} \frac{\nu_{aq}}{\nu_{sol}} 
\end{equation}
is the volume of reactive fluid needed to dissolve a solid volume equal to the initial pore space in the fracture. 
Fig.~\ref{porevolumes} shows contour plots of $V_{inj}/V_0$ in the $Pe-Da$ plane. A comparison with
the dissolution patterns in Fig.~\ref{AllPen} suggests that optimal injection rates ($Pe \approx 10-100, Da > 1$) do indeed correspond to a regime of strong channeling.

The values of $Pe$ and $Da$ cannot be varied independently in the same system, since both the diffusion constant and reaction rate are material properties. Changing the injection rate moves the system along a line of constant $PeDa=k \bar{h}_0 / D$, as shown by the dashed line in Fig.~\ref{porevolumes}. This produces a characteristic U-shaped 
dependence of $V_{inj}$ on Damk\"ohler number, an example of which is shown in the inset to Fig.~\ref{porevolumes}. 
The minimum in this curve corresponds to the optimal injection rate for a given value of $PeDa$.
A similar dependence of $V_{inj}$ on Damk\"ohler number has been reported previously~\citep{Fredd1998,Golfier2002,Panga2005,Kalia2007,Cohen2008}.

An important practical observation is that the optimum flow rate in the mass-transfer-limited regime apparently occurs at a constant value of the effective Damk\"ohler number~\citep{Golfier2002}. This result, based on Darcy-scale simulations of dissolution in two-dimensional porous media is consistent with the results in Fig.~\ref{porevolumes}. 
When $Da > 1$, the line of constant $Da_{eff} = 1/10$ (from Fig.~\ref{pdaeff}) runs near the valley of  minimal $V_{inj}(Pe,Da)$. However, when the reaction rate is reduced, the optimum path shifts to higher $Da_{eff}$, as can be seen in the inset to Fig.~\ref{porevolumes}.

\section{Wormhole competition and coarsening of the pattern}\label{sec:worms}

\begin{figure*}
\center\includegraphics[height=38pc, angle=-90]{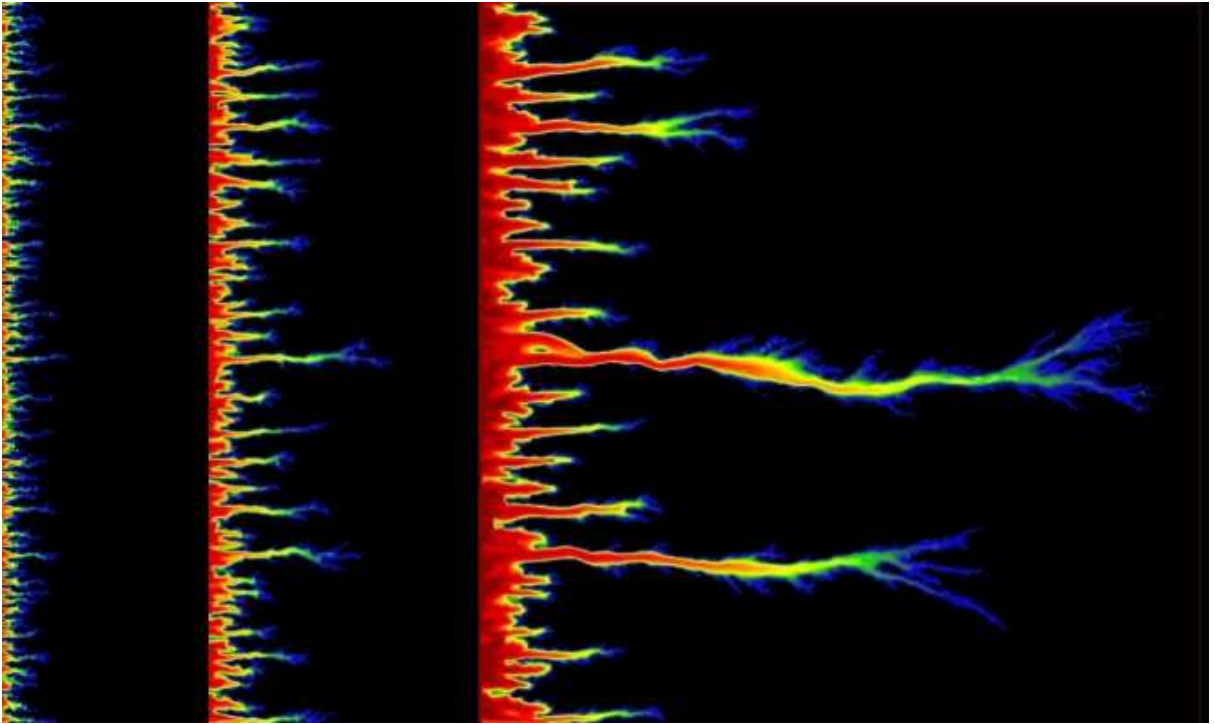}
\caption{(Color online) Erosion of the lower surface (initially flat) in an artificial
fracture at $Pe=32$ and $Da\rightarrow\infty$, captured at (from left to right) $\Delta \bar{h} =0.15 \bar{h}_0$, $0.5 \bar{h}_0$
and $2 \bar{h}_0$ respectively.}\label{dissol}
\end{figure*}

In Fig.~\ref{dissol}, the dissolution patterns for a larger fracture ($800 \times 800$ pixels) at $Pe=32$ and $Da\rightarrow\infty$ (constant pressure drop) are captured at three different instances, corresponding to an aperture increase $\Delta \bar{h} =0.15 \bar{h}_0$, $0.5 \bar{h}_0$ and $2 \bar{h}_0$ respectively.
Only a small fraction of the channels present at
$\Delta \bar{h}=0.15 \bar{h}_0$ persist to later times ($\Delta \bar{h} =0.5 \bar{h}_0$);
the channels that do survive have advanced far ahead of the dissolution front. The process of channel competition is self-similar,
and the characteristic length between active (growing) wormholes increases with time, while the number of active channels decreases; these systems have been shown to exhibit nontrivial scaling relations~\citep{Szy06}.
The competition between the emerging fingers leads to hierarchical structures that are characteristic of many unstable growth processes~\citep{Evertsz:90,Couder1990,Krug1997,Huang:1997,Gubiec2008}
from viscous fingering~\citep{Roy1999} and dendritic side-branches
growth in crystallization~\citep{Couder:2005} to crack propagation in brittle solids~\citep{Huang:1997}.
However, due to the finite size of the fracture system, the competition ends as soon as the fingers 
break through to the outlet. In fact, even before breakthrough the simple hierarchical growth pattern is disrupted, as shown in the rightmost panel of Fig.~\ref{dissol}. The large pressure gradient at the tips of the leading channels causes them to split into two or more daughter branches~\citep{Daccord1987,Hoe88,Fredd1998}. 

 \begin{figure}[t]
 \center\includegraphics[width=20pc]{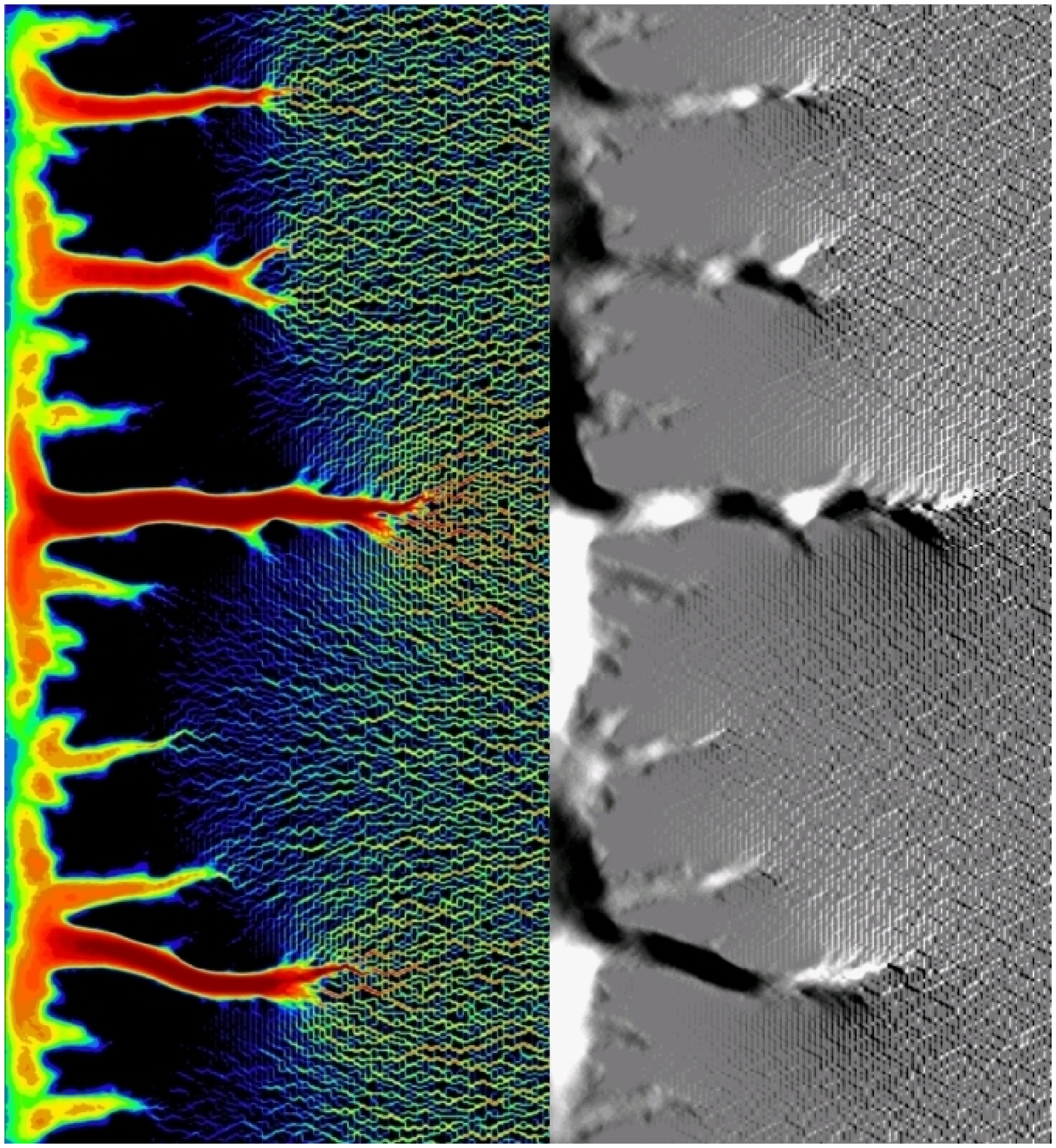}
 \caption{(Color online) Contours of the integrated flow field, $v^{2d} = \sqrt{{\bf v}^{2d} \cdot {\bf v}^{2d}}$, (left) and its lateral
component, $v_y^{2d}$, (right) in a small section of the fracture. In the left panel, the lightest/red shading indicates the regions of highest flow,
followed by green, blue, and black. In the right panel the shading corresponds to the {\em direction} of the lateral fluid velocity, white for $v_y^{2d} >0$ and black for  $v_y^{2d}<0$. The integrated flow field, ${\bf v}^{2d}$, is defined in Eq.~\eqref{v2d}.}\label{panor}
\end{figure}

\begin{figure}[t]
\center\includegraphics[width=30pc]{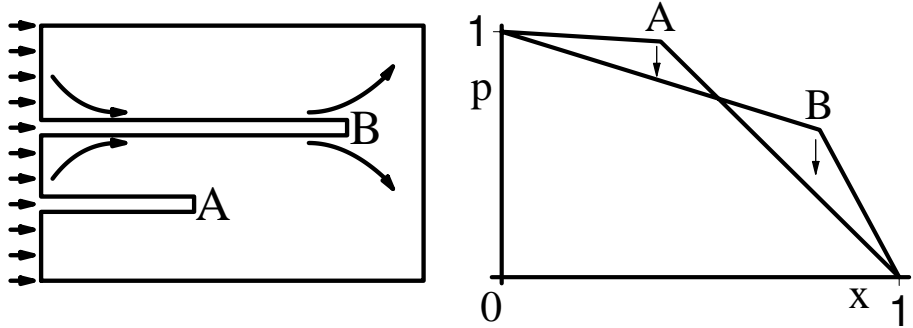}
\caption{Two dissolution channels in the fracture (left panel) and the
corresponding pressure drops (right panel). The flow lines are
converging towards a larger channel at the inlet and diverging
near the tip of the conduit. (This figure was taken from \citet{Szy06}).}\label{chann}
\end{figure}

The interaction between wormholes, which underlies the selection process, can be
investigated by analyzing the flow patterns in the dissolving fracture. Figure~\ref{panor}
shows a magnified view of a part of the sample containing just a few channels.
The left panel shows the magnitude of the fluid flow in the system, $v^{2d} = \sqrt{{\bf v}^{2d} \cdot {\bf v}^{2d}}$;
the flow is focused in a few active channels while the rest of the medium is
bypassed. The right panel shows the lateral ($v_y^{2d}$) component of the
flow: in the white regions $v_y^{2d}>0$ whereas in the black regions $v_y^{2d}<0$. It is observed that
the longest channel, positioned in the center of the sample, is draining flow from the two
channels above it as well as from the one underneath; $v_y^{2d}>0$ below the channel and $v_y^{2d}<0$ above it. Similarly, the 
second largest channel, situated in the lower part of the fracture is draining flow from the smaller channel immediately above it.
Thus, out of the four large channels in Fig.~\ref{panor}, only two are really active; the
other two just supply flow to the active channels.
A careful examination of the right panel of Fig.~\ref{panor} shows that at the downstream end of the channels
the flow pattern is reversed; the flow is now diverging from the active channels rather
than converging towards them as in the upstream part.

Fluid flow in the vicinity
of the channels is shown schematically in the left panel of Fig.~\ref{chann}, and can be understood from the
pressure drop in the channels~\citep{Szy06}, sketched in the right panel. With a constant total pressure drop between the
inlet and outlet, the pressure gradient in the long channel is steeper than in the short
channel, because the flow rate is higher. In the upstream part of the fracture
the short channel is therefore at a higher pressure than the long one, so flow in the surrounding matrix is directed towards the long
channel. Downstream the situation is reversed; the region around the tip of the long channel is at a higher pressure than the
surrounding medium and so flow is directed away from the channel, resulting in the converging-diverging flow pattern seen in the simulations. The larger the difference in channel lengths, the higher the pressure difference between the channels. The diverging flow pattern at the tips leads to the characteristic splitting seen near the breakthrough region in Fig.~\ref{dissol}, where the pressure gradients at the tips are large.

\begin{figure*}
\center\includegraphics[width=39pc]{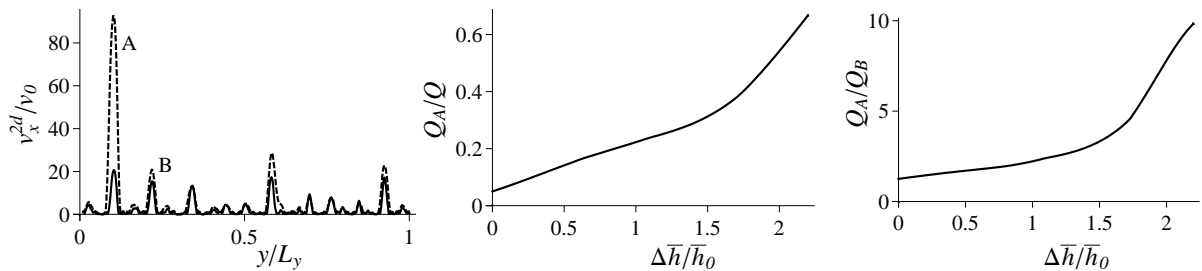}
\caption{Left: A cross-section of the flow map $v_x^{2d}(y)$ at $\Delta \bar{h} = \bar{h}_0$ (solid) and
$\Delta \bar{h} = 2 \bar{h}_0$ (dashed) ($Pe=8$ and $Da=0.1$). Only
four channels have increased their flow rates significantly in the corresponding time interval and at $\Delta \bar{h} = 2 \bar{h}_0$ about 55\% of the total flow is focused in the main channel.
Center: The fraction of flow focused in the main channel as a function of the dissolved volume $\Delta \bar{h}/\bar{h}_0$. The change of slope marks the point when the channel breaks through to the other side of
the fracture.
Right: The ratio of the flow focused in two neighboring channels (A and B) as a function
of $\Delta \bar{h}$}
\label{gx}
\end{figure*}

The higher mass flow rates in the longer channels lead to more rapid dissolution and
this positive feedback causes rapid growth of the longer channels
and starvation of the shorter ones. Channel competition is explicitly illustrated in
Fig.~\ref{gx}, which compares channel lengths and flow rates
at different stages of the dissolution process. At the beginning, about 20 channels were spontaneously
formed by the initial instability at the dissolution front, while at $\Delta \bar{h} \approx \bar{h}_0$ (left panel) only four of them
remain active. As dissolution progresses these remaining four channels self select, and at $\Delta \bar{h}=2 \bar{h}_0$ a single active channel transports more than a half of the total flow
through the sample (center panel). It can be seen that channel A which starts out only slightly longer than channel B, drains flow from channel B (right panel), slowly at first, but eventually the volumetric flow in channel A becomes an order of magnitude larger than that in B. This process repeats itself until only a single channel remains or breakthrough occurs.

\section{Influence of the initial topography on wormhole formation}\label{geom}

 \begin{figure*}
\center\includegraphics[width=39pc]{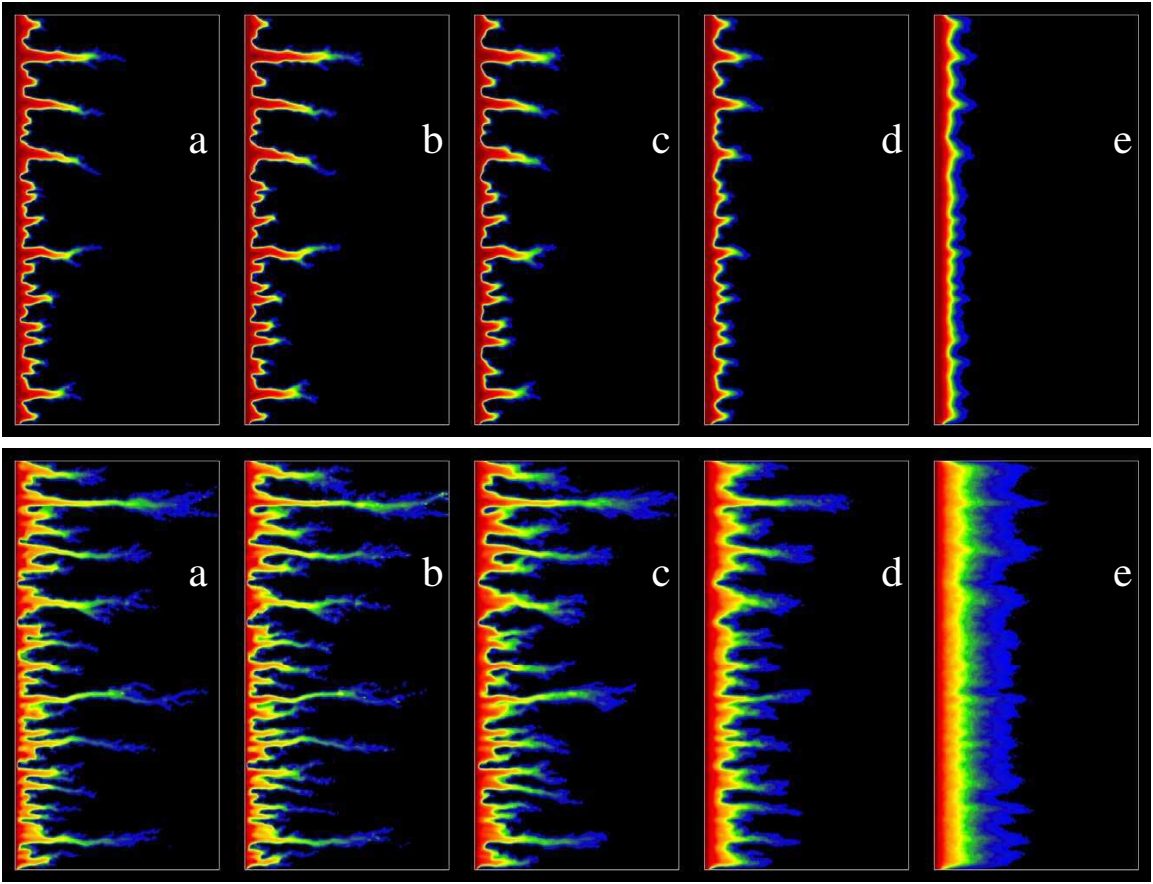}
 \caption{(Color online) Dissolution patterns for the original topography (a) and
for fractures with additional separation between the surfaces: $\hat{h}=0.1 \bar{h}_0$ (b)
$\hat{h}=0.5 \bar{h}_0$ (c), $\hat{h}=\bar{h}_0$ (d), and
$\hat{h}=2 \bar{h}_0$ (e). The flow rates (at constant pressure drop) correspond to initial P\'eclet numbers $Pe = 8$ (upper panel) and $Pe = 32$ (lower panel); $Da \rightarrow \infty$ in both cases. The erosion patterns of an initially flat surface were captured at $\Delta \bar{h} = \bar{h}_0$.}\label{shiftsPe08}
\end{figure*}

It was shown in Secs.~\ref{sec:channels} and~\ref{sec:perm} that initial topographies characterized
by a large number of contact
points ($\zeta=0.5$), relatively high roughness, and small mean aperture, have a well-defined
range of P\'eclet and Damk\"ohler numbers where spontaneous channeling is strong. However,
in the other geometric extreme, when fracture surfaces are far from each other and the roughness is
relatively small, the initial inhomogeneities in flow and aperture are smoothed out
as dissolution proceeds.
The effect of fracture roughness can be illustrated by comparing dissolution patterns computed
in the original fracture geometry, with those obtained after introducing an additional
separation (extra aperture) $\hat{h}$ between the surfaces, while
keeping the geometry ($\zeta = 0.5$), P\'eclet and Damk\"ohler numbers the same.
In this case the obstacles do not span the whole aperture, and the relative roughness is
a function of both $\zeta$ and $\hat{h}$,
\begin{equation}
f(\zeta, \hat{h}) = \left(\frac{\zeta (1-\zeta) {\bar h}_0^2}{\left[(1-\zeta){\bar h}_0+ {\hat h}\right]^2}\right)^{1/2}
\label{fzeta}
\end{equation}

Figure~\ref{shiftsPe08} shows the influence of additional aperture on the dissolution patterns for $Pe=8$ and
$Pe=32$, where the channeling was found to be strongest. We use an identical arrangement of obstacles
in each case ($\zeta = 0.5$), and the Damk\"ohler number $Da \rightarrow \infty$. The results show a rather well-defined threshold of additional aperture, $\hat h_{tr} \approx \bar{h}_0$,
beyond which wormholing is strongly suppressed. However, below the threshold the degree of channeling is scarcely affected, although the relative roughness is reduced from $f=1$ at $\hat{h} = 0$ to $f = 0.33$ at $\hat{h}=\bar{h}_0$. 
It was argued in \citet{Han98} that, at larger values of $f$, flow focusing and channel 
competition are strongly enhanced, whereas for small relative roughness the
flow is more diffuse and dissolution becomes uniform. Our results indicate a qualitative connection between the
roughness of the aperture and channel formation, but not a strong quantitative correlation. Despite the large reduction
in statistical roughness the dissolution patterns are essentially unchanged up to the threshold values of $\hat{h}$.
In contrast, for separations above the threshold, the patterns change dramatically,
with the  channels disappearing as the value of $\hat{h}$ is increased from approximately $ \bar{h}_0$ to $2 \bar{h}_0$, yet the statistical roughness decreases only slightly, from $f = 0.33$ to $f = 0.2$. The absence of a good quantitative correlation between $f$ and the degree of channel formation, coupled with the sharp transition to channel suppression suggests that other geometric factors, beyond statistical roughness, may play a significant role in determining wormholing. One such geometric factor is the degree of contact between the fracture surfaces, which will be examined below.

\begin{figure*}
\hspace{-0.5in}
\center\includegraphics[width=39pc]{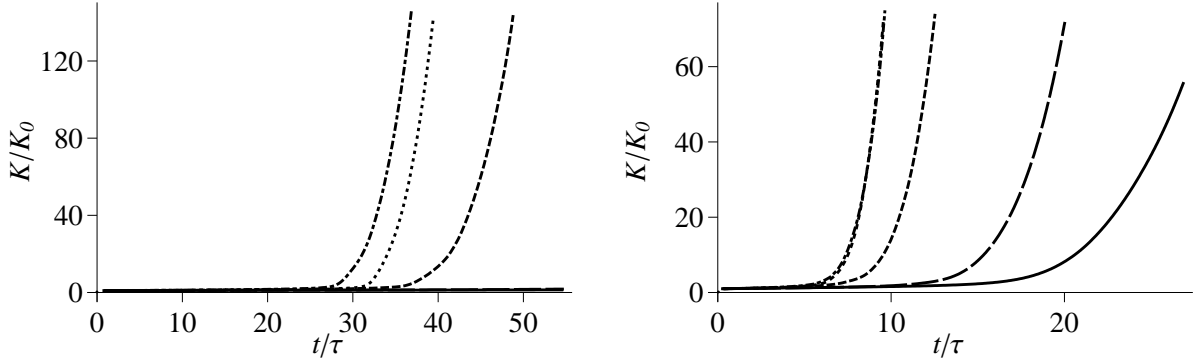}
\caption{Time evolution of the permeability
at $Pe=8$ (left) and $Pe=32$ (right). Results are shown for different additional apertures:
$\hat{h}=0$ (dotted), $\hat{h}=0.1 \bar{h}_0$ (dot-dashed), $\hat{h}=0.5 \bar{h}_0$ (dashed),
$\hat{h}= \bar{h}_0$ (long dashes) and  $\hat{h}=2 \bar{h}_0$ (solid). In the left panel the plots for $\hat{h}= \bar{h}_0$  and $\hat{h}=2 \bar{h}_0$
overlap and lie almost along the horizontal axis, while in the right panel the plots for $\hat{h}=0$  and $\hat{h}=0.1 \bar{h}_0$
overlap.}\label{shiftflows}
\end{figure*}

The evolution of permeability, Fig.~\ref{shiftflows}, shows
that a small amount of additional separation between the surfaces ($\hat{h} = 0.1 \bar{h}_0$)
does not suppress wormholing ($Pe=32$) and can even speed up both dissolution
and channel formation ($Pe=8$).
This illustrates two important points. First, a small gap between the fracture surfaces blocks
the flow pathways almost as effectively as complete contact. Second, a small
additional separation enhances the flow rates in the spontaneously formed channels,
which leads to faster growth, despite the reduction in roughness.

\begin{figure*}
\center\includegraphics[width=30pc]{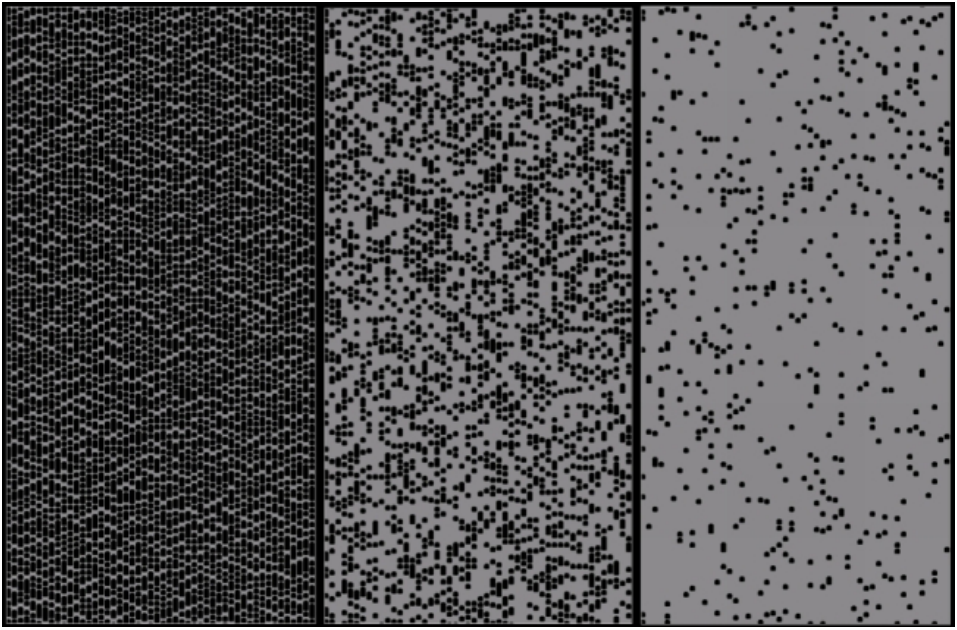}
\caption{Initial fracture geometries at coverages $\zeta = 0.5$ (left),
$\zeta = 0.25$ (center), and  $\zeta = 0.05$ (right).}\label{geomdil}
\end{figure*}

\begin{figure*}
\center\includegraphics[width=39pc]{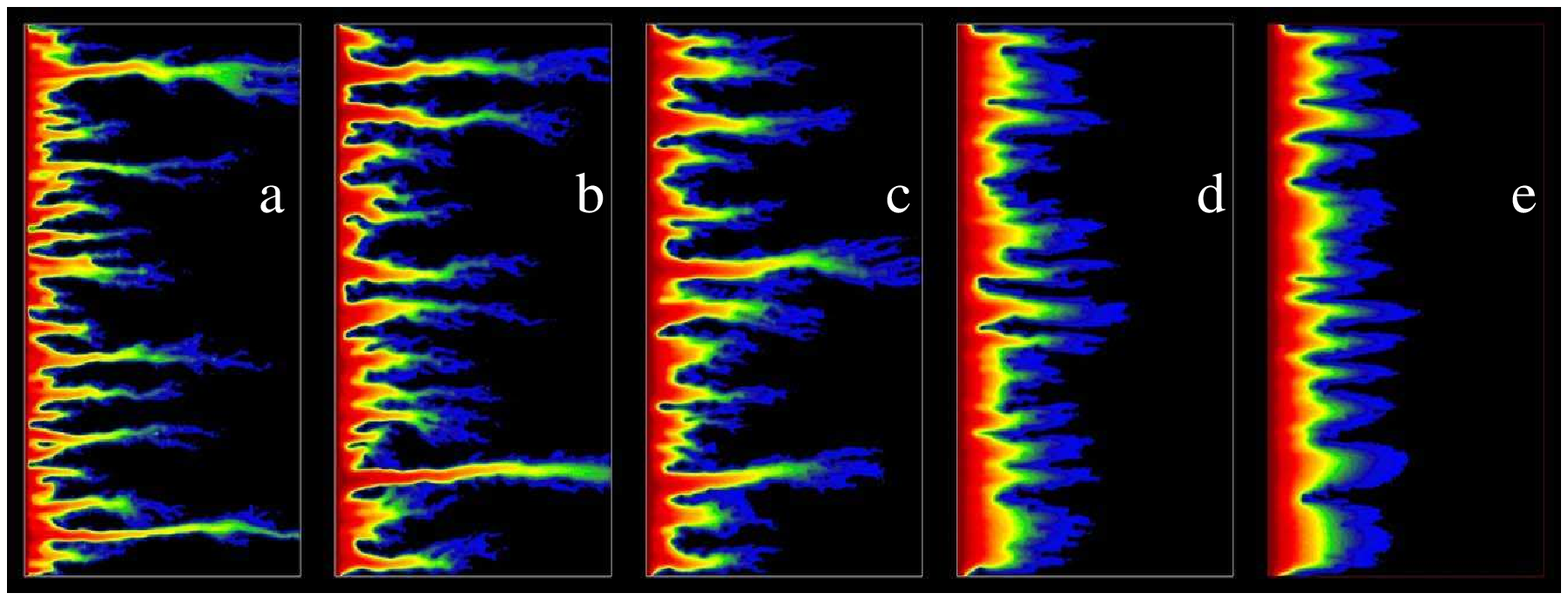}
\caption{(Color online) Dissolution patterns for different coverages (a) $\zeta = 0.5$, (b)  $\zeta = 0.25$, (c)  $\zeta = 0.15$,  (d) $\zeta = 0.05$, and (e) $\zeta = 0.025$. The erosion of an initially flat surface at $Pe=32$, $Da\rightarrow\infty$, was captured at $\Delta \bar{h} = 1.25 \bar{h}_0$. 
}\label{dilutionsPe32h1}
\end{figure*}

Further insight into the influence of geometry on the dissolution patterns may be gained
by reducing the initial coverage, $\zeta$, randomly removing some of the protrusions from the original fracture, as illustrated
in Fig.~\ref{geomdil}. The roughness is again reduced, but in a geometrically different fashion from that represented in Fig.~\ref{shiftflows}.
Figure~\ref{dilutionsPe32h1} shows the resulting dissolution patterns at $Pe = 32$, $Da\rightarrow\infty$.  Again, there seems to be a well defined threshold where channeling is suppressed, occurring at coverages between $\zeta=0.05$ and $\zeta=0.1$, which, according to Eq.~\eqref{fzeta} corresponds to a relative roughness $f \approx 0.2 - 0.3$. Above the threshold, both the diameter of the channels and the spacing between them are
weakly dependent on the geometry of the system. This supports the notion
that the characteristics of wormhole formation are primarily functions of $Pe$ and $Da$, and essentially independent of 
the correlation length of the fracture topography.

\section{Summary}

In this paper we have studied channeling instabilities in a single fracture, using
a fully three-dimensional, microscopic numerical method. Channeling was found to be 
strongest for large reaction rates (mass-transfer-limited regime) and intermediate P\'eclet numbers. We analyzed the observed patterns in terms of two simple models; a Darcy-scale model for the reaction front and a convection-diffusion model for mass transport in the wormhole. These models qualitatively explain the wide variety of dissolution patterns observed in the simulations, and we found quantitative agreement in the prediction of the wormhole diameter in the mass-transfer-limited regime. We summarized our simulations in terms of a phase diagram separating the different regimes of erosion, and compared our conclusions to experiments and other numerical simulations. We also examined  the efficiency with which permeability can be increased by acid erosion.

When wormholes form, there is strong competition for the flow, leading to an attrition of the shorter channels. We have previously explained a number of detailed observations by a simple network work of the flow in the fracture system \citep{Szy06}. In this paper we have provide new simulation data showing explicitly how the fluid flow is drained from the matrix surrounding the dominant channels. This provides confirmation at the microscopic level for key assumptions underpinning the network model.

We found evidence for the existence of a well-defined threshold value of the fracture roughness, $f\approx 0.2-0.3$, needed to trigger a channeling instability. Below that value, the topographic perturbations are smeared out by dissolution on similar or shorter timescales than the propagation of the front, and 
channels do not develop. Above the roughness threshold channels do develop, but their size and spacing are controlled by the values of P\'eclet and Damk\"ohler number, and not by the fracture topography.

Our results are limited to geometries characterized by short-range spatial correlations and lacking
the self-affine properties of natural fractures. The analysis of wormholing in such geometries will be the subject of future study.

\begin{acknowledgments}
This work was supported by the Polish Ministry of Science and Higher Education
(Grant No. N202023 32/0702), and by the US Department of Energy, Chemical Sciences, Geosciences and Biosciences Division, Office of Basic Energy Sciences (DE-FG02-98ER14853). We would like to refer the interested reader to a recent paper~\citep{Kalia2009}, published after the present manuscript was submitted, that is relevant to the analysis of Sec.~\ref{geom}.
\end{acknowledgments}


\end{document}